\documentclass[%
aps,
prd,
twocolumn,
superscriptaddress,
preprintnumbers,
nofootinbib,
amsmath,amssymb,
]{revtex4-2}

\usepackage{style}
\usepackage{aas_macros}
\usepackage[para]{threeparttable}

\begin{document}

\title{Precision Solar System Dynamics for Ultralight Dark Matter Search}

\author{Jonas Frerick}
\affiliation{Dipartimento di Fisica, Sapienza Universit\`a di Roma \& Sezione INFN Roma1, Piazzale Aldo Moro 5, 00185, Roma, Italy}

\author{Hyungjin Kim}
\affiliation{Universit\'e Paris-Saclay, CNRS, CEA, Institut de Physique Th\'eorique, 91191 Gif-sur-Yvette, France}

\author{Felix Kling}
\affiliation{Universit\"at Bonn, Regina-Pacis-Weg 3, D-53113 Bonn, Germany}

\begin{abstract}
Ultralight dark matter exhibits an order-one density fluctuation at the scale of its wavelength. This density fluctuation exists across the entire dark matter halo and interacts with stars and planets, perturbing their motion via gravitational interactions. We investigate the possibility of using precision solar system dynamics to search for ultralight dark matter. We examine this possibility with interplanetary radio range measurements. We show that the precision of current range measurements can probe ultralight dark matter at masses around $10^{-15} \eV$, had its density in the solar system been $10^5$ larger than the so-called local dark matter density. This limit complements other constraints, such as the one from analyses of pulsar timing observations. 
\end{abstract}

\maketitle
\tableofcontents

\bigskip

\section{Introduction}
Ultralight dark matter (ULDM) remains a compelling dark matter candidate. It refers to a light bosonic dark matter candidate below the eV scale. Due to its bosonic nature and minuscule mass, it behaves like classical waves, leading to interesting phenomenological consequences throughout the history of the universe on various scales. See recent reviews on the subject~\cite{Niemeyer:2019aqm, Hui:2021tkt, OHare:2024nmr, Eberhardt:2025caq} for details.

Ultralight dark matter exhibits an order-one density fluctuation in the dark matter halo. This order-one density fluctuation may perturb the motion of ordinary matter in our universe. When this effect is integrated over astrophysical time scales, it may change the orbital evolution of massive objects, such as supermassive black holes, globular clusters, or may perturb the structure of star clusters~\cite{Hui:2016ltb, Marsh:2018zyw, Bar-Or:2018pxz, Bar:2021jff, Dalal:2022rmp, Teodori:2025rul}. These considerations constrain ultralight dark matter at extremely small masses, $m \lesssim 10^{-20}\eV$.

The order-one ULDM fluctuation perturbs not only stars but also any matter that gravitates. It is therefore conceivable that one might use gravitational wave detectors to measure ULDM-induced density fluctuations, as those detectors are designed to measure minuscule fluctuations in the positions of test masses. This idea has been examined in the recent literature using interferometric gravitational wave detectors, pulsar timing arrays, and astrometry observations~\cite{Kim:2023pkx, Kim:2023kyy, Kim:2024xcr, Eberhardt:2024ocm, Dror:2025nvg}. In particular, with the 12.5 year dataset of the NANOGrav pulsar timing collaboration~\cite{NANOGrav:2020gpb, hyungjin_2024_10534322}, the previous work obtained a limit on the ULDM density around the solar system as $\rho < 3\times 10^3\rho_0$ at $m=10^{-17}\,{\rm eV}$ where $\rho_0=0.4\,\GeV/{\rm cm}^3$ is the local dark matter density~\cite{Kim:2023kyy}. Current and future astrometry missions exhibit similar sensitivities around the same mass range~\cite{Kim:2024xcr}. Additionally, coherent density fluctuations provide another way to probe ULDM gravitationally with terrestrial detectors and observations~\cite{Khmelnitsky:2013lxt, Aoki:2016kwl, Porayko:2018sfa, NANOGrav:2023hvm, EuropeanPulsarTimingArray:2023egv, Sakstein:2023hvw, Yu:2024enm, Badurina:2025xwl}. In this work, we focus on the stochastic fluctuations, as it will allow us to probe a less-explored dark matter mass range around $m\sim 10^{-15}\eV$; searching for coherent density fluctuation will probe $m\lesssim 10^{-21}\eV$, a mass region which has been already probed by cosmological and astrophysical observations~\cite{Hui:2021tkt}.

The interpretation of pulsar timing and astrometry measurements requires accurate planetary ephemerides. As we---the observer---are moving around the solar system, a precise knowledge of the location of the observer as well as the solar system barycenter is required in order not to introduce a large systematic uncertainty in pulsar timing and astrometry. This leads to the intriguing question of whether one can directly use precision measurements of the solar system that are used to construct planetary ephemerides for ultralight dark matter searches. 

\begin{figure*}[t]
\centering
\includegraphics[width=0.7\textwidth]{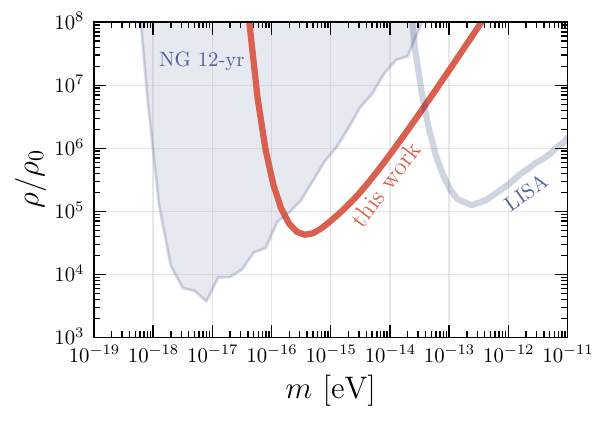}
\caption{
The projected sensitivity from a series of residual range measurements on the dark matter density near the solar system (red). We also show the excluded region from the null result of the analysis of NANOGrav 12.5 year pulsar timing dataset~\cite{NANOGrav:2020gpb, hyungjin_2024_10534322}, and the projected sensitivity of the future gravitational wave detector LISA~\cite{Gan:2026vph}. For the plot, we assume $T= 20\,{\rm yr}$, and $N=5$ independent range measurements. We also assume that the noise is white and is given by $S_n(f) = 2 \sigma_L^2 \Delta t =(300\,{\rm m/\sqrt{Hz}})^2$. Here $\rho_0=0.4 \, \GeV/{\rm cm}^3$ is the local dark matter density. A more detailed explanation is given in Section~\ref{sec:Projection}. 
}
\label{fig:projection}
\end{figure*}

The purpose of this work is therefore to address this question and examine the sensitivity of such measurements. Some work has already been done in this direction, focusing on local overdensities centered around the Sun. Pitjev and Pitjeva examined the motion of planetary perihelia to limit a uniform and spherically symmetric additional matter component within a certain orbital radius inside the solar system~\cite{Pitjev:2013sfa}. It was found that at Saturn's orbit, the density of such additional matter must satisfy $\rho / \rho_0 < 2 \times 10^4 $. The other works include the analysis of tracking of an asteroid by Tsai et al~\cite{Tsai:2022jnv}, which placed an upper limit on dark matter density as $\rho/\rho_0 < 6\times 10^6$, and the analysis of LAGEOS geodetic satellite and lunar laser ranging by Adler~\cite{Adler:2008rq}, which placed an upper limit of $\rho/\rho_0 < 10^{11}$ for Earth-bound dark matter. These limits apply when the dark matter is bound to the Sun or the Earth. This could happen for ultralight dark matter bound to the solar system via self-interaction~\cite{Budker:2023sex}, although we will not be concerned about this scenario in the present work.
For similar use of precision solar system orbit data to constrain new physics, see also~\cite{Tsai:2021irw,Tsai:2023zza, KumarPoddar:2020kdz, Aliberti:2024udm}.

A particularly interesting and related work has been done recently by Foster et al~\cite{Foster:2025csl}. The authors analyzed the evolution of orbital elements of a binary system due to ULDM coherent and stochastic density fluctuations. Using binary pulsar systems and the Earth-Moon system with lunar laser ranging measurements, it was found that an ultralight dark matter density around the solar system of $\rho / \rho_0 \gtrsim 10^{10} $ can be probed at mass $m \simeq 10^{-15}\eV$. We will compare their result with ours in more detail in Section~\ref{sec:Discussion}.

Building upon previous works, we study the implications of precision solar system measurements for stochastic density fluctuations of ultralight dark matter. In particular, we consider interplanetary range measurements that have been used to construct the planetary ephemerides. This data consists of the measured distances between Earth and other planets, for instance, Mercury, Venus, Mars, Saturn, and Jupiter. Uncertainties in these measurements are currently at the ${\cal O}(1)\,{\rm m}$ scale~\cite{2021AJ....161..105P, 2021NSTIM.110.....F}. We investigate the sensitivity of the range measurements on the ultralight dark matter density around the solar system by examining the ULDM-induced signal in the auto- and cross-correlation of such measurements. The result is already summarized in Fig.~\ref{fig:projection}, which will be detailed in subsequent sections. The results show that interplanetary range measurements could probe new parameter space of ultralight dark matter with peak sensitivity at $m\simeq 10^{-15}\,$eV, complementing the existing search from pulsar timing observations and future space-borne gravitational wave detectors LISA.

This work is organized as follows. In Section~\ref{sec:ULDM}, we review the basic properties of the ultralight dark matter field and its density fluctuations. Furthermore, we provide some intuitive discussion for the length scales relevant to the problem in the so-called quasiparticle picture. In Section~\ref{sec:SolarSystem}, we discuss how a precision model of the solar system is constructed from various measurements. We then identify the interplanetary distance measurements as observables for our study and compute the ULDM signal in those observables. In Section~\ref{sec:Projection}, we review the cross-correlation statistics. We compute the correlation of interplanetary distances and the resulting signal-to-noise ratio of the statistics to project the sensitivity of a collection of range measurements. In Section~\ref{sec:Discussion}, we discuss the effects of the acceleration of the Sun, potential sources of the correlated noise, and the current and future range measurements that could be used for ULDM searches. In Section~\ref{sec:Conclusion}, we conclude by summarizing the original question that we attempted to address in this work and potential future directions. In the Appendix, we provide further details on the analytic computation of the signal spectrum, numerical simulation, and a systematic estimation of the cross-correlation. We use natural units with $c = \hbar = 1$ throughout this work. 

\section{Ultralight Dark Matter}\label{sec:ULDM}
We review the basic properties of the underlying ultralight dark matter field in the solar neighborhood. We consider a minimally coupled scalar field whose action is given by
\begin{align}
S = \int d^4x \, \sqrt{-g} 
\left[
- \frac{1}{2} g^{\mu\nu} \partial_\mu \phi \partial_\nu \phi
- \frac{1}{2} m^2 \phi^2\;,
\right]
\end{align}
where $g_{\mu\nu}$ is a metric tensor, and $m$ is the mass of the dark matter. We ignore potential self-interactions between dark matter and non-gravitational interactions with the Standard Model particles. We choose the conformal Newtonian gauge to parameterize the metric, 
\begin{align}
ds^2 = -  ( 1 + 2 \Phi ) dt^2 + ( 1 - 2 \Psi ) dx^2\;,
\end{align}
where $\Phi$ and $\Psi$ are scalar potentials. Since we focus on the stochastic density fluctuations, we do not distinguish between these potential and approximate $\Psi \approx \Phi$. This can be justified by the small anisotropic stress induced by a non-relativistic DM background.

We assume that the dark matter field can be described by a Gaussian random field. We expand the field as
\begin{align}
\phi(t, \boldsymbol x)
\approx \frac{1}{\sqrt{2mV}} 
\sum_{\boldsymbol k}
\left[
a_{\boldsymbol k}e^{ i k \cdot x}
+ a_{\boldsymbol k}^* e^{- i k \cdot x}
\right]\;,
\end{align}
in the non-relativistic limit. This is a Fourier expansion of the field in a finite box of volume $V$. Here $\boldsymbol k$ is the three momentum. We treat $(a_{\boldsymbol k}, a_{\boldsymbol k}^*)$ as classical random variables. When decomposed into the amplitude and phase, $a_{\boldsymbol k} = r_{\boldsymbol k} e^{i \theta_{\boldsymbol k}}$, each quantity is distributed according to the Rayleigh and uniform distribution, respectively, with the following probability density function~\cite{Derevianko:2016vpm, Foster:2017hbq, Centers:2019dyn, Kim:2021yyo, Cheong:2024ose}
\begin{equation}
p(r_{\boldsymbol k})
= 
\frac{2 r_{\boldsymbol k}}{f_{\boldsymbol k}}
\exp\left[
- \frac{r_{\boldsymbol k}^2}{f_{\boldsymbol k}}
\right]
\quad \text{and} \quad
p(\theta_{\boldsymbol k}) 
= \frac{1}{2\pi}  \;.
\label{eq:pdf_random}
\end{equation}
In the continuum limit, the Rayleigh distribution parameter $f_{\boldsymbol k}$ can be understood as the dark matter momentum distribution, normalized as $\bar\rho = m \int [d^3k/(2\pi)^3] f(\boldsymbol k)$ with $\bar\rho$ being the mean dark matter density.

The quantity of primary importance is the correlation function of the fluctuation in the dark matter density. The density fluctuation is defined as
\begin{align}
\delta \rho(t, \boldsymbol x) = \rho(t, \boldsymbol x) - \bar \rho \;,
\end{align}
where the density operator is 
\begin{align}
\rho(t, \boldsymbol x)
= \frac{1}{2}
\left[
\dot\phi^2
+ (\nabla\phi)^2
+ m^2\phi^2
\right]\;. 
\end{align}
The two-point correlation function of the density fluctuation in Fourier space is
\begin{align}\label{eq:dens_corr}
\langle 
\widetilde{\delta \rho}(k)
\widetilde{\delta \rho}^*(k')
\rangle
= (2\pi)^4 \delta^{(4)}(k - k') P_{\delta \rho}(k)\;,
\end{align}
where $P_{\delta\rho}(k)$ is the power spectrum of the density fluctuation, which depends on the four-momentum $k=(\omega, {\boldsymbol k})$. It was pointed out in Ref.~\cite{Kim:2023pkx} that the power spectrum consists of two  sub-components, each centered at two distinctive frequencies; one at $\omega \simeq 2m$ and the other at $\omega \lesssim m \sigma^2$ where $\sigma = 160\,{\rm km/sec}$ is the velocity dispersion of dark matter.
They are referred to as coherent and stochastic density fluctuations, respectively. In this work, we focus on stochastic density fluctuations where the power spectrum is given by~\cite{Kim:2023kyy}
\begin{align}\label{eq:dens_spec}
P_{\delta\rho}(k)
= \frac{2\pi^2 \bar\rho^2}{m^4 \sigma^5} \frac{\sigma}{v_k}
\exp\left[
- \frac{v_k^2}{4\sigma^2}
- \frac{(\omega\tau)^2}{(v_k/\sigma)^2}
\right]\;,
\end{align}
where $v_{k} = |\boldsymbol{k}| /m$ and $\tau = 1/ m\sigma^2$. To derive this expression, we assume a normal velocity distribution with zero mean. A non-vanishing mean velocity due to the solar system motion introduces directionality in the system. This directionality could provide additional means to distinguish an ultralight dark matter signal from other astrophysical sources in case of positive signals at the price of computational complication. However, in the following, we ignore the mean velocity for the sake of simplicity. 

An intuitive way to understand stochastic density fluctuations is to consider them as if they are quasiparticles of size and mass~\cite{Hui:2016ltb}
\begin{align}
\lambda_{\rm DM}
&= \frac{1}{m\sigma} 
= 2.5\, {\rm AU}
\times 
\Big(
\frac{10^{-15}\,{\rm eV}}{m}
\Big) \;, 
\label{quasi_size}
\\
M_{\rm eff}
&= \bar \rho \lambda_{\rm DM}^3
= 4\times 10^{13} \,{\rm kg} 
\times
\Big(
\frac{10^{-15}\,{\rm eV}}{m}
\Big)^3 \;,
\label{quasi_mass}
\end{align}
where we use $\bar\rho = \rho_0$ for the estimates. This suggests that the quasiparticle could be as large as a few astronomical units (AU) and as massive as small asteroids in the solar system. Unlike other massive objects in our solar system, this quasiparticle is a transient object; it appears and disappears in a stochastic manner in a time scale given by the coherence time $\tau =  1 / m \sigma^2$, which has also been observed in numerous numerical simulations e.g. Refs.~\cite{Schive:2014dra, Schive:2014hza}. These massive quasiparticles continuously perturb the motion of the solar system, potentially leaving stochastic signals in range measurements. This picture will allow us to imagine how the solar system dynamics might be used for an ultralight dark matter search. 

\section{Solar System Ephemeris}\label{sec:SolarSystem}
Precision modeling of the solar system has played an important role in fundamental science. This includes deep space navigation, planetary science, and gravitational wave searches with pulsar timing, among many others. Planetary ephemerides are generated by fitting numerically integrated planetary orbits to various observations, such as interplanetary range, lunar laser range, Doppler, very long baseline interferometery (VLBI), and astrometry measurements. Range measurements mainly determine the relative distance between planets, while VLBI and astrometry measurements determine the sky location of the planets. The result of the fitting is the planetary ephemeris and the residuals in respective measurements. The residual in range measurements is currently at a few-meter scale for planetary range measurements, while the residual angle in sky location of planets remains at the sub-microarcsecond level~\cite{2021NSTIM.110.....F, 2021AJ....161..105P}.

We examine the possibility of using two-way planetary range measurements for an ultralight dark matter search. In two-way range measurements, a round-trip light travel time between a ground station and a satellite is measured. From such measurements, we estimate the one-way range as
\begin{align}
L = \frac{\Delta \tau}{2} \;,
\end{align}
where $\Delta \tau$ is the round-trip light travel time. The stochastic density fluctuation of dark matter could leave signals in this observable by (i) perturbing the motion of planetary objects, (ii) inducing a time delay of radio signals, and (iii) introducing gravitational redshift in the observer's location. These effects can be schematically written as 
\begin{align}
\Delta \tau (t)= 
(\Delta \tau)_{\rm O}(t)
+ (\Delta \tau)_{\rm S}(t)
+ (\Delta \tau)_{\rm E}(t)\;,
\end{align}
where $({\Delta \tau})_{\rm O}$ is due to the orbital perturbation, $(\Delta\tau)_{\rm S}$ represents the Shapiro delay, and $(\Delta \tau)_{\rm E}$ denotes the Einstein time delay, which is due to the discrepancy between the coordinate time and observer's proper time. Each of them is given by
\begin{align}
\!\! \!\! 
(\Delta \tau)_{\rm O}
&= 
\hat{\boldsymbol n}_a \cdot 
\big[ ( \boldsymbol x_a(t_s) - \boldsymbol x_r(t) ) 
+  ( \boldsymbol x_a(t_s) - \boldsymbol x_r(t_t) ) \big]
\label{tau_O}\;,
\\
\!\! \!\! 
(\Delta \tau)_{\rm S}
&=
- 2 \omega_\gamma 
\bigg( 
    \int^{\lambda}_{\lambda_s} d\lambda' 
    + \int^{\lambda_s}_{\lambda_t} d\lambda 
\bigg) \Phi
\label{tau_S}\;,
\\
\!\! \!\! 
(\Delta \tau)_{\rm E}
&= \int^t_{t_t} dt \, \bigg( \Phi - \frac{1}{2} v^2 \bigg)
\label{tau_E}\;,
\end{align}
where $t_t \approx t - 2L$ and $t_s \approx t - L$ are the coordinate time at the transmission and at the arrival to the satellite, $\hat{\boldsymbol n}_a$ is the unit vector from the reference to the satellite $a$, $\boldsymbol x_{a,r}$ is the position of the satellite $a$ and the reference observer, $\lambda$ is an affine parameter describing the worldline of photon with which the photon four-momentum is defined $p^\mu = dx^\mu /d \lambda$, $\omega_\gamma$ is the frequency of radio wave, and $v$ is the velocity of the observer on Earth. The integral in the Shapiro and Einstein delay signal should be understood as an integral over the worldline of the photon and the observer, respectively. 

In the following, we focus on the orbital perturbation signal, $(\Delta\tau)_{\rm O}$. Without detailed computation, we can estimate the relative importance of each effect. By treating the density fluctuation as a quasiparticle, Eqs.~\eqref{quasi_size} -- \eqref{quasi_mass}, an acceleration and potential induced by a quasiparticle can be estimated as
$$
a_q \sim \frac{G M_{\rm eff}}{\lambda_{\rm DM}^2}
\quad\textrm{and}\quad
\Phi_q \sim \frac{G M_{\rm eff}}{\lambda_{\rm DM}}\;.
$$
With the above approximation, we find
\begin{align}
(\Delta \tau)_O 
&\sim a_q \tau_{\rm DM}^2 
\sim
1\,{\rm cm}
\Big( \frac{10^{-15}\eV}{m}\Big)^3\; , 
\nonumber
\\
(\Delta \tau)_S 
&\sim \Phi_q \lambda_{\rm DM} \Big(\frac{L}{\lambda_{\rm DM}}\Big)^{\frac12}
\sim 3\,{\rm nm} 
\Big( \frac{10^{-15}\eV}{m} \Big)^{\frac52} \Big( \frac{L}{3\,{\rm AU}} \Big)^{\frac12}\; , 
\nonumber
\\
(\Delta \tau)_E 
&\sim \Phi_q L 
\sim 3\,{\rm nm} 
\Big( \frac{10^{-15} \eV}{m} \Big)^2 \Big( \frac{L}{3\,{\rm AU}} \Big)\; , 
\nonumber
\end{align} 
over one coherence time scale with $\bar \rho / \rho_0= 10^5$. Each estimation allows an intuitive understanding. The first line is a typical displacement of test masses over $\tau_{\rm DM}$ due to the quasiparticle acceleration $a_q$. The second line is the Shapiro delay of light, Eq.~\eqref{tau_S}, through $N_{\rm patch}=(L/\lambda_{\rm DM})$ incoherent patches of gravitational potential in the limit of $\lambda_{\rm DM} < L$. The third line is the Einstein delay, Eq.~\eqref{tau_E}, assuming that the potential does not change over the round-trip light travel time of ${\cal O}(1)$ minutes. This estimation already supports our treatment of considering only the orbital perturbation signal. We justify this treatment in the Appendix~\ref{app:signal} more carefully through detailed computations of the signal power spectra.  

\subsection{Stationary Range Response}\label{sec:stationary}
We now illustrate the response of the observable with respect to the ultralight dark matter fluctuations. For this purpose, we will consider the one-way range due to the orbital perturbation
\begin{align}
\Delta L_a(t) = 
\hat{\boldsymbol n}_a \cdot \big[ \boldsymbol x_a(t - \bar L_{a}) - \boldsymbol x_{r}(t) \big] - \bar L_{a}\;,
\end{align}
where $\boldsymbol x_{a, r}(t)$ is the position vector of the test mass $a$ and the reference mass $r$, $\hat{\boldsymbol n}_a$ is the nominal unit vector from the reference to the test mass $a$, and $\bar L$ is the nominal separation of the two test masses without dark matter perturbation. We choose to work with the one-way range since the response is approximately identical to that of the two-way range for the parameters under consideration. Each displacement vector evolves according to the equation of motion,
\begin{align}\label{eq:eoms}
\ddot{\boldsymbol x}_{i}
= \boldsymbol{a}_{\rm ss}(t, \boldsymbol x_i) - \nabla \Phi(t, \boldsymbol x_i)\;,
\end{align}
where $\boldsymbol{a}_{\rm ss}(t, \boldsymbol x)$ is the acceleration on each test mass due to the solar system objects, and $\nabla \Phi(t,\boldsymbol x)$ is the force due to the stochastic ULDM fluctuation. The potential is linked to the underlying density fluctuation via the Poisson equation
\begin{align}
\nabla^2 \Phi = 4 \pi G \delta \rho\;. 
\end{align}
The goal of the rest of this section is to characterize the dark matter signal power spectrum of the residual range measurement. 

The external force tends to complicate the computation by rendering the system nonlinear in its displacement, even in the limit where the perturbation is small. To proceed, we will ignore the external force for a moment and compute the signal as if the test and reference masses were free particles, subject only to ULDM-induced fluctuations. This approximation could be justified in the limit where the length of the time series is much smaller than a typical orbital time scale of solar system objects. This point will be revisited in the discussion section and justified both analytically and numerically.

With this simplification, the computation can be done using the method introduced in the previous section. The signal power spectrum is defined via the two-point function of the residual range:
\begin{align}\label{eq:correlator_pre}
\big\langle
\widetilde{\Delta L}_a(f)
\widetilde{\Delta L}_b^*(f')
\big\rangle
= \frac{\delta(f - f')}{2} 
\Sigma_{ab}(f, \bar{\boldsymbol L}_a, \bar{\boldsymbol L}_b)\;,
\end{align}
where the Fourier component is defined as $\widetilde{\Delta L}_a(f) = \int dt \, e^{2\pi i f t} \Delta L_a(t)$. We decompose the cross-spectral density as
\begin{align}\label{eq:correlator}
\Sigma_{ab}(f, \bar{\boldsymbol L}_a, \bar{\boldsymbol L}_b)
= S_x^{\rm DM}(f) \,
{\cal I}_{ab}(f, \bar{\boldsymbol{L}}_a, \bar{\boldsymbol{L}}_b)\;. 
\end{align}
Here $\bar{\boldsymbol L}_{a}$ is the nominal displacement vector from the reference to the test mass $a$ without dark matter perturbations.
The spectrum $S^{\rm DM}_x(f)$ is the signal spectrum of a single test mass position projected onto an arbitrary direction, $\hat{\boldsymbol n}$, and is given by
\begin{align}
S_x^{\rm DM}(f) 
&=
\frac{2}{(2\pi f)^4}
\int \frac{d^3k}{(2\pi)^3} 
\Big( \frac{4\pi G}{k^2} \Big)^2 P_{\delta\rho}(k)
(\boldsymbol k \cdot \hat{\boldsymbol n})^2
\nonumber\\
&= \frac{2}{(2\pi f)^4} \frac{(4 \pi G \rho)^2}{3 m^3 \sigma^4} K_0(2\pi f \tau)\; ,
\label{single_testmass_spectrum}
\end{align}
where $K_0(x)$ is a modified Bessel function.  The function ${\cal I}_{ab}(f, \bar{\boldsymbol L}_a, \bar{\boldsymbol L}_b)$ encodes information on the response of the range measurement as a function of the separation, frequency, and dark matter parameters. A detailed computation of the spectrum and response integral is given in Appendix~\ref{app:Computation_Spectrum}. 

To understand the general behavior of the response, we quote our results of the auto-correlation ($a=b$) in two limits from the appendix: (i) the short wavelength limit, $\lambda_{\rm DM} \ll L$, and (ii) the long wavelength limit, $\lambda_{\rm DM} \gg L$. In the short wavelength limit, we find ${\cal I}_{aa} \sim 2$, indicating that the perturbation in the position of each mass due to ULDM is uncorrelated, and hence the fluctuation in the range is simply an incoherent sum of fluctuations in each test mass.\footnote{The short wavelength limit has a subtle frequency dependence. Nonetheless, the response integral takes ${\cal O}(1)$ values for all frequencies, see Appendix~\ref{app:Computation_Spectrum} for a detailed discussion.} In the long wavelength limit, we find ${\cal I}_{aa} \propto (L_a/\lambda_{\rm DM})^2$, which reflects the tidal suppression of the residual range as both masses experience approximately the same force from the ULDM density fluctuation.

For later discussion, we introduce the correlation coefficients of two range measurements. We define it as
\begin{align}
\Gamma_{ab}(f, \bar{\boldsymbol L}_a, \bar{\boldsymbol L}_b) = \frac{{\cal I}_{ab}(f, \bar{\boldsymbol L}_a, \bar{\boldsymbol L}_b)}{\sqrt{{\cal I}_{aa}(f, \bar{\boldsymbol L}_a, \bar{\boldsymbol L}_a) {\cal I}_{bb}(f, \bar{\boldsymbol L}_b, \bar{\boldsymbol L}_b) }} \;. 
\label{corr_coef}
\end{align}
The definition is identical to the Pearson correlation coefficient. It is normalized as $\Gamma_{aa}=1$, and describes the degree of correlation between two independent range measurements. This contains unique information on the correlation of the ULDM-induced signal as a function of the nominal ranges $\bar L_{a,b}$, frequency $f$, and the angular separation of two different test masses parametrized as $\hat{\boldsymbol n}_a \cdot \hat{\boldsymbol n}_b$. 

\subsection{Non-Stationary Range Response}
We assumed in the previous section that the nominal separation and the angular separation are constant. Although this might be true over a time scale shorter than a typical orbital time scale, the values of these quantities eventually drift slowly as the test and reference masses evolve over a longer time scale. This implies that the cross-spectral density $\Sigma_{ab}(f, \bar{\boldsymbol L}_a, \bar{\boldsymbol L}_b)$ and and the correlation coefficient $\Gamma_{ab}(f, \bar{\boldsymbol L}_a, \bar{\boldsymbol L}_b)$ depend explicitly on time.

\begin{figure}
\centering
\includegraphics[width=0.45\textwidth]{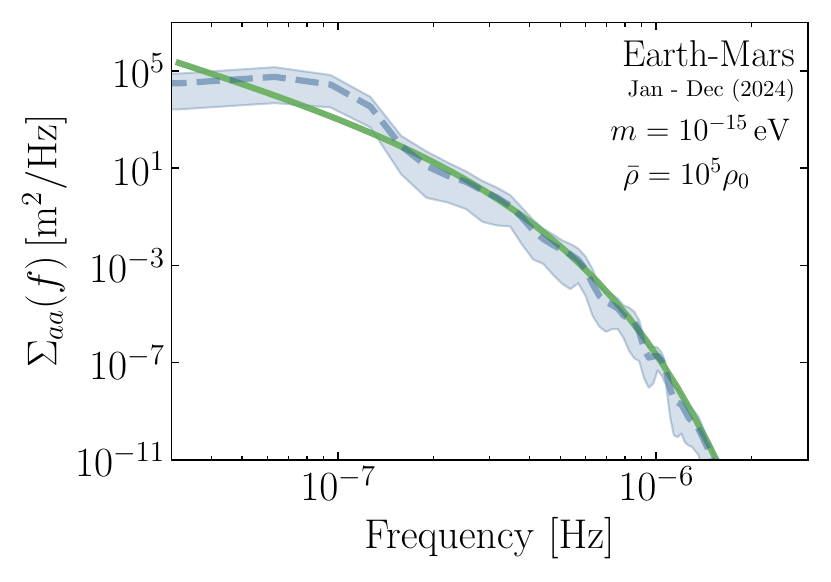}
\caption{The auto-correlation of the Earth-Mars residual range due to ultralight dark matter. We use the orbits of Earth and Mars during the year 2024. The solid green line is the analytically computed mean auto-correlation during the same time interval. For numerical estimation, we ran $N=10^2$ simulations. From each simulation, we obtain the residual range and estimate the auto-correlation. The dashed line is the mean value of the estimated auto-correlation from the ensemble of simulations, while the lower and upper boundary of the band are the $5\%$ and $95\%$ quantile, respectively. For this result, we choose $m=10^{-15}\eV$ and $\bar \rho / \rho_0 = 10^5$ for the dark matter, and $n=4$ for the order of the polynomial detrending, as well as $\alpha=6$ for the windowing of the residual data. See the main text for details.}
\label{fig:spectrum_Mars}
\end{figure}

We still expect that the analytic results on the cross-spectral density in the previous section faithfully describe the fluctuations in the range measurement for the time scale shorter than a typical orbital time period. This is expected because the separation and the angle between each observation remain approximately constant over such a time period. Nonetheless, it is still important to check to what degree this expectation is correct. For this, we run an ensemble of numerical simulations of the solar system with ultralight dark matter. For each numerical simulation, we obtain time series of residual ranges, segment them into smaller pieces, each with the total duration of a few months, and numerically estimate the power and cross-spectral densities. We investigate if (i) numerically obtained spectra agree with the analytic result in Eq.~\eqref{eq:correlator}, and (ii) the correlation coefficient estimated from simulations agrees with the analytic result in Eq.~\eqref{corr_coef}.

\begin{figure*}[!t]
\centering
\includegraphics[width=0.95\textwidth]{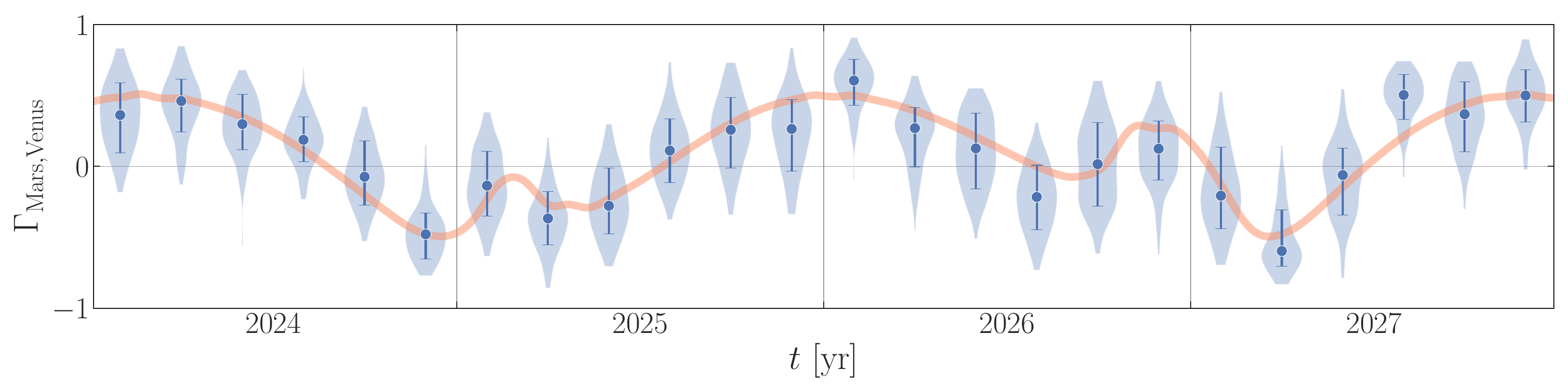}
\caption{The correlation coefficient between two range measurements, Earth-Mars and Earth-Venus, from the same set of numerical simulations. The solid curve is obtained from the analytic result, with the nominal displacement during the years 2024 -- 2027 as an input. In each time interval, the auto-correlation and cross-spectral density are estimated, and the correlation coefficient is derived. The mean value of the numerical result agrees well with the analytic one. For this plot, we fixed the frequency to $f= 7\times 10^{-6}\,{\rm Hz}$ and chose $m=10^{-14}\eV$ and $\bar \rho/\rho_0 = 10^{10}$. 
}
\label{fig:spectrum_time}
\end{figure*}

We first verify the auto-correlation of the residual range. The numerical simulation includes the innermost six planets with a post-Newtonian correction to the Keplerian equation. The implementation is detailed  in Appendix~\ref{app:Numerical_Simulation}. From each simulation, we obtain the range between Earth and Mars as a function of time. To calculate the residual range, we obtain two sets of distances: one with dark matter and the other without dark matter. We then compute the difference between the two time series to obtain the residual range. The resulting residual is further detrended by polynomials of degree $n$ to account for a systematic drift due to the ULDM density fluctuation and the gravitational potential of the Sun.\footnote{We emphasize that this drift is part of the ULDM signal. In fact, Ref.~\cite{Foster:2025csl} investigated the long term evolution of binary systems in this context. However, in the context of our formalism we discard this additional component of the signal from our analysis. The combination of both signal components is expected to yield even stronger bounds.} We then apply the following window function to the residual range data to taper the spectral leakage:
\begin{align}\label{window}
w(t) = \sin^\alpha(\pi t / T)\;,
\end{align}
where $\alpha$ is a positive integer.  The resulting power spectrum is shown in Fig.~\ref{fig:spectrum_Mars}. For this result, we choose $m= 10^{-15}\eV$ and $\bar \rho / \rho_0 = 10^5$ for dark matter, $\alpha=6$ for the window function, and $n=4$ for the polynomial detrending. This result uses the trajectories of Earth and Mars during the year 2024. The numerical result agrees with the analytic estimation, with deviations at the lowest and highest frequencies shown in the figure. The lowest frequency corresponds to $f_{\rm yr} = 1/{\rm yr} = 3 \times 10^{-8}\,{\rm Hz}$ where we do not expect our analytic result to hold. In addition, an application of the window function and polynomial fitting partially removes dark matter signal and distorts the spectrum at low frequencies. 
The auto-correlation of other residual ranges is given in the appendix. 

The computation of the signal spectrum and the correlation are carried out by assuming that the nominal displacement $\boldsymbol L_{a,b}$ is constant. As was already mentioned earlier, these nominal displacements evolve over the time scale of a year. This time dependence might be captured simply by replacing the response integral with
\begin{equation}\label{eq:stat_to_dyn}
{\cal I}_{ab}(f, \bar{\boldsymbol L}_a, \bar{\boldsymbol L}_b)
\to 
{\cal I}_{ab} \big(f, \bar{\boldsymbol L}_a(t), \bar{\boldsymbol L}_b(t) \big)
\end{equation}
provided that we estimate the spectral densities with a time series shorter than a year.

To verify the time-dependence of the cross-correlation, we estimate the correlation coefficient in Eq.~\eqref{corr_coef} numerically from our simulation. We consider the time span 2024 -- 2027 and run simulations of the solar system to obtain two range measurements: one between Earth and Mars and another between Earth and Venus. The residual range of each measurement is computed similarly by taking the difference of the distances between a target planet and Earth with and without dark matter. Using them, we estimate both the auto- and cross-spectral density of the two resulting residual range via the Welch method~\cite{1967ITAE...15...70W}, and compute the correlation coefficient. In Fig.~\ref{fig:spectrum_time}, we show the correlation coefficient as a function of time during the chosen time period. The numerical result agrees with the analytic result, verifying that the computation given in the previous section is valid for frequencies above an inverse of a year.

\section{Projection}\label{sec:Projection}
Having verified the validity of our analytic description of the cross-spectral density, we adopt the optimal cross-correlation statistic to estimate the sensitivity of range measurements for ultralight dark matter. The optimal cross-correlation statistic was previously introduced for stochastic gravitational wave searches with a network of ground-based gravitational wave detectors~\cite{Allen:1997ad}. The statistic is designed to estimate the cross-correlation of independent data streams that is induced by a stochastic gravitational wave. Since our dark matter signal is also stochastic and expected to leave correlated signals in range measurements, the optimal cross-correlation statistic can be similarly used for our dark matter search. See Refs.~\cite{Maggiore:2007ulw, Romano:2016dpx} for a pedagogical introduction. 

We consider a data stream
$$
d_a(t) = s_a(t) + n_a(t)\;,
$$
where the subscript $a=1,\,2,\, \cdots, N$ runs over different range measurements, $d_a(t)$ is a measured residual range, $s_a(t) = \Delta L_a(t)$ is an ULDM-induced signal, and $n_a(t)$ is a zero-mean Gaussian random noise. The signal is characterized by the two-point function [cf.\ Eq.~\eqref{eq:correlator}], while the noise is fully characterized by
\begin{align}
\langle
\widetilde n_a(f)
\widetilde n_b^*(f')
\rangle
= \frac{1}{2} \delta(f - f') \delta_{ab} S_{n_a}(f) \;,
\label{noise_correlator}
\end{align}
where $S_{n_a}(f)$ is the one-sided noise power spectrum. In writing the noise power spectrum, we assume that the noise in each range measurement is uncorrelated. We discuss the validity of this assumption in the next section. 

The time span of the data stream is usually longer than a typical orbital time scale. In such a case, we segment it into smaller intervals,
\begin{align}
d_a(t) = \{ d_a^{(I)}(t) \,\, | \,\, I = 1 , \, 2 ,\, \cdots ,\, N_{\rm seg} \}\;,
\end{align}
where the time span of each segment $d_a^{(I)}(t)$ is small enough so that the nominal separation $\bar{\boldsymbol L}_a(t)$ remains approximately constant. With a segmented time series, we construct a cross-correlation statistic segment-by-segment,
\begin{align}
\hat Y_{ab}^{(I)} 
= \int_{-\infty}^\infty df \, 
\tilde d_a^{(I)*}(f)
\tilde d_b^{(I)}(f)
\widetilde Q^{(I)}_{ab}(f)\;,
\end{align}
where $\widetilde Q^{(I)}_{ab}(f)$ is an arbitrary filter. The signal and noise are defined by the mean and variance in the presence and absence of dark matter signal, $\mu_{ab}^{(I)} = {\mathbb E}[\hat Y_{ab}^{(I)}]$ and $(\sigma_{ab}^{(I)})^2 = {\rm Var}(\hat Y_{ab}^{(I)})$, respectively. The optimal filter that maximizes the signal-to-noise ratio $\mu_{ab}^{(I)} / \sigma_{ab}^{(I)}$ is given by~\cite{Allen:1997ad, Maggiore:2007ulw} 
\begin{align}
\widetilde Q^{(I)}_{ab}(f)
= \frac{\Sigma_{ab}\big(f, \bar{\boldsymbol L}_a^{(I)}, \bar{\boldsymbol L}_b^{(I)}\big)}{S_{n_a}(f) S_{n_b}(f)}\;. 
\end{align}
Here $\bar{\boldsymbol L}_{a,b}^{(I)}$ is a representative value of the nominal separation in this segment $I$.

The statistic $Y_{ab}^{(I)}$ is then combined together with a weighting factor,
\begin{align}
\hat Y = \sum_{a> b, I} w_{ab}^{(I)} \hat Y_{ab}^{(I)}\;. 
\end{align}
The weighting factor is chosen to maximize the signal-to-noise ratio of the statistic $\hat Y$. It is straightforward to find the optimal weighting factor as $w_{ab}^{(I)} = \mu_{ab}^{(I)}/ \sigma_{ab}^{(I)2}$. In the weak signal limit where the variance of the statistic is dominated by the noise, the signal-to-noise ratio is given by~\cite{Allen:1997ad} 
\begin{align}\label{eq:snr}
{\rm SNR}
= 
\bigg[
\sum_I 2 T^{(I)} 
\sum_{a> b} 
\int_{f_\ell}^{f_u} df\, 
\frac{|\Sigma_{ab}(f, \bar{\boldsymbol L}_a^{(I)}, \bar{\boldsymbol L}_b^{(I)})|^2}{S_{n_a}(f) S_{n_b}(f)}
\bigg]^{\frac12}\;,
\end{align}
where $T^{(I)}$ is the time span of the segment $I$, $f_\ell= 1/T^{(I)}$, $f_u = 1 / (2 \Delta t)$, and $\Delta t$ is the cadence of the measurements. We assume that the statistic for each segment is independent. 

We show in Fig.~\ref{fig:projection} the projected sensitivity based on the signal-to-noise ratio with range measurements. We consider a set of $N=5$ baseline range measurements over a $T=20$ year time span with a cadence of $\Delta t =1\,{\rm day}$ as a benchmark. We envision that these five range measurements are between Earth and 5 other planets, namely Mercury, Venus, Mars, Saturn, and Jupiter. We assume that the noise in each measurement is white and universal; its spectrum is given by $S_{n_a}(f) = S_{n_b}(f) = 2\sigma_L^2 \Delta t = [300 \, {\rm m}/\sqrt{\rm Hz}]^{2}$, where $\sigma_L$ is the rms error of the residual range measurement and $\Delta t$ is the cadence of observation. Although the noise of each residual range is generally different from each other, we nonetheless expect that the above choice is reasonable as $\sigma_L = {\cal O}(1)\,{\rm m}$ and $\Delta t =1\,{\rm day}$ for most range measurements. This will be further justified in Section~\ref{sec:Discussion}. The length of each segment is assumed to be $T^{(I)} = 6\,{\rm months}$. Since the signal-to-noise ratio scales as ${\rm SNR} \propto (\bar \rho / \sigma_L)^2$, the projected sensitivity will improve linearly with the rms noise of the range measurement. Although we choose the universal noise power spectrum, it is conceivable that one of the measurements might provide a dominant sensitivity at a given mass since the response integral depends on the relative size of the nominal separation and the dark matter wavelength. 

The projection in Fig.~\ref{fig:projection} shows that the range measurements could probe the dark matter density in the solar system gravitationally, had its density been $10^{5}$ larger than the so-called local dark matter density for the mass range around $10^{-15}\eV$. Below this mass range, the sensitivity quickly drops as the stochastic signal appears below the lowest possible frequency accessible with the segmented time series, i.e. at $f_{\rm sig} \lesssim m \sigma^2 / 2\pi < f_\ell$. On the other hand, above $10^{-15}\eV$, the wavelength of dark matter becomes smaller than a few AU scale. With the approximate expression of the correlator in the short-wavelength limit, we expect the sensitivity to scale as $\bar \rho \propto m^{3/2}$, which is consistent with the result shown in the figure. For this figure, we do not use accurate orbital trajectories of planets, but assume that they are on a circular orbit with their average semi-major axis for simplicity. 

\section{Discussion}\label{sec:Discussion}

\subsection{Effects of Acceleration due to the Sun}\label{sec:sol_acc}

The signal power spectrum and the resulting projections are based on the response of residual range measurement, assuming that the test masses are free particles. This treatment is valid only for  time series sufficiently shorter than the typical orbital time scale of a year. For this reason, we envision an analysis that segments the time series of the range measurements into smaller pieces of $T_s < 1\,$year. This limits the lowest possible frequency that can be resolved to $f_\ell = 1 / T_s$, effectively setting a minimal DM mass to be probed. As the signal power spectrum scales as $1/f^4$, this segmentation limits the sensitivity of the range measurement for ULDM search. 

One might attempt to use a time series that spans more than the dynamical time scale of the solar system. In this case, the system described by Eq.~\eqref{eq:eoms} becomes non-linear due to the acceleration of solar system objects. The effect of such an acceleration cannot be modeled as a drift in the position and velocity of the planets as before. Deriving an analytic expression for the signal power spectrum amounts to solving non-linear stochastic differential equations, which we were unable to do.\footnote{See, however, the recent discussions on the evolutions of orbital parameters based on kinetic equation in binary systems \cite{Blas:2021mpc, Blas:2021mqw, Caputo:2026fuj}.} We nonetheless expect that the gravitational force in the equation of motion will taper the spectrum at frequencies below the characteristic frequency of the orbital motion for the reasons described in the following. 

Consider a simple one-dimensional harmonic oscillator subjected to a Gaussian white noise $\xi(t)$. In this case, the equation of motion is given by
$$
\ddot{x}
= -\omega_0^2 x + \xi(t)\;,
$$
where $\omega_0$ is the characteristic frequency of the system. The spectrum of the position of the harmonic oscillator is then given by
$$
S_x(\omega) = \frac{ S_\xi(\omega) }{(\omega^2 - \omega_0^2)^2}\;,
$$
where $S_\xi(f) = {\rm const}$ is the power spectrum of the Gaussian random noise $\xi(t)$. From the above expression, we observe that the spectrum behaves as $S_x(\omega) \propto 1 /\omega^4$ for $\omega > \omega_0$, similar to Eq.~\eqref{single_testmass_spectrum}, and that it saturates to a constant value $S_x(\omega) \sim S_\xi/\omega_0^4$ at frequencies lower than the characteristic frequency $\omega_0$. 

Unlike a simple harmonic oscillator, the equation of motion for planets is non-linear in its position, and therefore, it is non-trivial how the above argument might apply. Nevertheless, one may linearize the equation of motion around an unperturbed nominal trajectory, and find that the linearized equation has a similar structure as the simple harmonic oscillator with a driving frequency given by the inverse of a typical orbital period of planets. This again suggests that the analytic computation is valid when $T_s < 1\,{\rm year}$, which is why we have chosen to divide the time series into multiple segments with a duration of a few months. This critical frequency above which we trust the analytic expression also depends on the target planet. For example, the numerically reconstructed spectrum of the Earth-Mercury residual range agrees with the analytic one when $T_s \lesssim 6\,{\rm months}$ rather than a year. This is confirmed with full numerical simulations; see discussions in the Appendix~\ref{app:Numerical_Simulation}.

\subsection{Sources of Correlated Noise}
We have assumed uncorrelated noise for the sensitivity projection. When correlated noise exists, it could introduce a bias in the cross-correlation statistic, thereby rendering the discrimination of the dark matter signal from other systematics difficult in the presence of any positive signals. Although we do not expect the projected sensitivities to change significantly even when the noise is correlated, we nonetheless investigate potential sources of correlated noise in the range measurements.

The correlated noise is parameterized by the following two-point function of noise:
$$
\langle \tilde n_a(f) \tilde n_b(f') \rangle
= \frac{1}{2} \delta(f - f') S(f) \Gamma_{ab}\;,
$$
where $\Gamma_{ab}$ parametrizes the correlation. The correlation coefficient $\Gamma_{ab}$ is normalized as $\Gamma_{aa}=1$. Depending on the origin of the noise, the correlation coefficient might take different forms. 

Atomic clocks provide a source of correlated noise. In the two-way range measurements, the round-trip light travel time $\Delta \tau_a$ is measured with the frequency standard of the ground station, which is given by hydrogen masers~\cite{DSN304}. Any uncertainty in the frequency standard will appear in the range observable as correlated noise with a monopole-like pattern $\Gamma_{ab}^{\rm clock} = 1$. When the frequency standard fluctuates, $f \to f + \delta f$, the time delay measured by the standard fluctuates as $\Delta \tau_a \to \Delta \tau_a ( 1 + \delta  f / f)$, and thus it leads to a fluctuation in the measured range, $\Delta L_a = \bar L_a (\delta f /f)$. With a few hundred seconds of averaging time, hydrogen maser clocks in ground stations achieve a fractional frequency uncertainty of $10^{-14}$, which translates into $\Delta L_a \sim {\cal O}(1)\,{\rm mm}$ for $\bar L_a \sim {\rm AU}$. Compared to the effect expected from dark matter and the noise that we took as a benchmark, this is negligible. This can be translated into the following noise spectrum:
$$
S^{\rm clock}(f) = \bar L_a \bar L_b S_{\delta f / f}(f)\;,
$$
where $S_{\delta f /f}(f)$ is the spectrum of fractional frequency uncertainty. The routinely available performance of a local hydrogen maser is roughly consistent to the flicker noise, $S_{\delta f /f}(f) = (10^{-28}/ {\rm Hz}) ({\rm Hz}/f)$ for an averaging time $ \tau \in [10^2, \, 10^5]\,{\rm sec}$~\cite{DSN304}. By comparing the benchmark noise $S_n(f) = (300\,{\rm m} /\sqrt{\rm Hz})^2$ and an extrapolated frequency uncertainty spectrum $S_{\delta f /f}(f)$ to $f = 10^{-7}\,{\rm Hz}$, we observe that clock noise is subdominant throughout the frequency range of our analysis. Therefore, we do not expect the uncertainty of the local frequency standard to significantly affect the projected sensitivity or to induce a significant bias in the statistic. 

Time-tagging uncertainties provide another source of correlated noise. The master clock of each Deep Space Network station is synchronized with a United States Naval Observatory realization of the coordinate universal time, UTC(USNO). Any offset $\Delta t$ between two time standards results in a correlated error of
$$
\Delta L = \dot{L} \Delta t\;. 
$$
The time offest between two standards is maintained at the level of $< 3 \mu{\rm sec}$~\cite{DSN304}. With $\dot L \sim 1\,{\rm AU}/{\rm yr}$, one finds the time-offset error as $\Delta L \sim {\cal O}(1)\,{\rm mm}$. Furthermore, the offset between UTC and UTC(USNO) remains at a few nanoseconds level~\cite{BIPM_database}. The time-tagging error is therefore irrelevant for the purpose of our analysis. 

Unidentified objects in the solar system could also induce correlated noise via long-range gravitational interactions. Fedderke et al examined the implications of asteroids on the performance of a potential network of gravitational wave detectors in the solar system at frequencies below $10^{-7} \,{\rm Hz}$~\cite{Fedderke:2020yfy}. For a hypothetical setup consisting of two test masses, each with an orbital radius and a separation of $1\,{\rm AU}$, the noise due to objects in asteroid belts was estimated as $S_{a}^{\text{ast}}(f) \sim 10^{-22} \,[{\rm (m /s^2)^2/Hz}]$ when the 50 most massive objects in the asteroid belts are modeled and subtracted from the data stream. The differential acceleration power spectrum due to ultralight dark matter is around $S_{a}^{\text{ULDM}}(f) \sim 10^{-22} [{\rm (m /s^2)^2/Hz}]$ in the same frequency range with parameters at which we achieve our peak sensitivity, i.e., $m=10^{-15}\eV$, and $\rho/\rho_0 = 10^5$, so they turn out to be of the same order. 
 
We expect this noise estimate to be conservative for our analysis. When constructing the planetary ephemeris, around 300 asteroids---which have the most significant impacts on Mars' orbit and which constitute $90\%$ of the total mass of the asteroid belt---were identified and integrated iteratively to obtain the planetary ephemeris~\cite{2013Icar..222..243K, 2014IPNPR.196C...1F, 2021AJ....161..105P}. The resulting Mars residual range does not seem to exhibit a characteristic behavior of acceleration noise in the time domain from the time scale of a day to a year (e.g. Fig.~7 in Park et al~\cite{2021AJ....161..105P}).  Even when we use the estimate given by Fedderke et al~\cite{Fedderke:2020yfy} to estimate displacement noise $S_{x}(f) = S_a(f) / (2\pi f)^4$, the magnitude of the resulting noise power spectrum at the lowest possible frequency in our analysis $f = 1/{\rm yr}$ remains smaller than the white noise power spectrum we adopted in our study. In addition, we expect that the correlation pattern would be different as the spatial morphology of the asteroid and ultralight dark matter is different. For this reason, we expect that perturbations from the asteroid belt would not significantly change our conclusion. 

Another relevant source of correlated noise is the frequency-dependent delay of the range signal due to solar plasma. This effect varies according to the signal path with respect to the Sun and the solar cycle. This noise from the solar plasma is identified as one of the limitations of the current range measurements~\cite{doi:https://doi.org/10.1002/0471728454.ch3, 2018AGUFM.P41F3791P}. As the noise arises from the interplanetary plasma, it is conceivable that it might correlate some of the different range measurements. The level of bias will depend on the correlation length and coherent time of the solar plasma, and examination of them requires a careful modeling of the solar plasma, which is beyond the scope of this study. While this bias affects the parameter estimation in the event of positive detection of signal, we do not expect that it affects out limits significantly as we choose the noise power spectrum such that it reflects the current limitation. See Ref.~\cite{doi:https://doi.org/10.1002/0471728454.ch3} for a more detailed discussion of such noise sources.

\subsection{Status of Ranging Measurements}

\begin{table*}[!t]
\renewcommand{\arraystretch}{1.2}
\setlength\tabcolsep{8.pt}
\centering
\begin{tabular}{l | lllll}
\toprule
Mission
& Duration\tnote{b}

& Target
& Cadence [day]
& rms [m]
& ASD\tnote{a} [$\rm m/\sqrt{\rm Hz}$]

\\ \hline 
MESSENGER 
& $2011 - 2016$

& Mercury
& $1$
& $0.7$
& $3\times 10^2$
\\
BepiColombo
& $2027-$
& Mercury
& $1$
& $10^{-2}$
& $4$
\\
Venus Express
& $2006-2014$
& Venus
& $1$
& $8$
& $4\times 10^3$
\\
Mars Express
& $2005 - $
& Mars
& $1$
& $2$
& $7\times 10^2$
\\
Mars Odyssey
& $2002 - $
& Mars
& $1$
& $0.7$
& $2\times 10^2$
\\
Mars Reconn. Orbiter
& $2006 - $
& Mars
& $1$
& $0.7$
& $3 \times 10^2$
\\
Mars Global Surveyor
& $1999 - 2007$
& Mars
& $1$
& $0.7$
& $3\times 10^2$
\\
Cassini
& $2004 - 2017$
& Saturn
& $30$
& $3$
& $7\times 10^3$
\\
Juno
& $2016 - $
& Jupiter
& $90$
& $13$
& $5\times 10^4$ 
\\
\bottomrule
\end{tabular}

\caption{Summary of each ranging mission.  The rms error is taken from Ref.~\cite{2021AJ....161..105P}. The cadence of observations of the missions in the inner planets are around $1$ day, while it is $\sim30$ days for Cassini and $\sim90$ days for Juno. This can be estimated from partially available data in the website of NASA Solar System Dynamics as of 21 May 2026, and the data provided in Table 3 -- 5 of Ref.~\cite{2021AJ....161..105P}. The amplitude spectral density is then inferred from the estimated cadence and rms range error, assuming that the noise is white. The start and end dates represent the beginning and conclusion of the mission. The duration represents the span of range observation; for the ended missions, the values are taken from Ref.~\cite{2021AJ....161..105P}, while for the missions extended beyond the year 2025, only the starting date is quoted from the same reference. For BepiColombo, we use the value quoted in Ref.~\cite{2018AGUFM.P41F3791P} for the rms error, and assume a similar cadence of observations as in other missions in the inner solar system. 
}
\label{tab:noise_estimate}
\end{table*}

Range measurements have played a crucial role in space navigation over the last few decades. Currently, the performance of such measurements is at the level of a few-meter scale. This level of sensitivity allows not only for space exploration, but also to carry out research on fundamental science, e.g. testing general relativity and modified Newtonian dynamics~\cite{Bertotti:2003rm, Hees:2014kta}, gravitational waves~\cite{Fedderke:2021kuy, Blas:2021mqw}, and dark matter~\cite{Tran:2023jci, Zwick:2024hag, Foster:2025nzf, Foster:2025csl}. The status of some missions in the past two decades is summarized in Table~\ref{tab:noise_estimate}. As mentioned earlier, the root-mean-square error is at the level of ${\cal O}(1)$ m, which corresponds to a noise amplitude spectral density (ASD) of ${\cal O}(10^2) \, {\rm m/\sqrt{\rm Hz}}$ assuming that the noise is white and that the cadence of observation is one day. 

The precision of range measurements is expected to improve by an order of magnitude in future missions. The performance of the interplanetary range has remained at the level of a few meters for the last three decades, and the main source of noise has been the time delay due to the solar plasma and station delay within the ground station~\cite{2018AGUFM.P41F3791P}. Future missions such as BepiColombo aim to improve the accuracy by implementing a dual-frequency radio link system to subtract the solar plasma delay and by improving station delay calibration. The expected accuracy of such advanced range instruments is at the level of ${\cal O}(1\textrm{--}10)$ cm~\cite{2018AGUFM.P41F3791P, Iess2021}, which is 1--2 orders of magnitude better than the accuracy of the current range instruments. Given that the signal-to-noise ratio scales as ${\rm SNR} \propto (\bar \rho / \sigma_L)^2$, this improvement will allow the sensitivity of the measurements on dark matter density to improve by the same factor.

\subsection{Comparison to Other Works}
To place our results in the context of existing studies of gravitational probes of ultralight dark matter, we compare our results with the results obtained in other works. We particularly focus on the comparison between ours and the one using lunar laser ranging~\cite{Foster:2025csl}, pulsar timing and astrometry~\cite{Kim:2023kyy, Kim:2024xcr}.

Foster et al examined the binary dynamics between the Earth and the Moon. Using the stochastic density fluctuation of dark matter, they estimated the sensitivity of lunar laser ranging for ultralight dark matter as $\rho/\rho_0 \lesssim 10^{10}$ at $m=10^{-15}\eV$~\cite{Foster:2025csl}. This differs from the result presented in this work, cf.\ Fig.~\ref{fig:projection}. This difference may be attributed to the difference in the choice of the system---the lunar laser range versus the interplanetary radio range---and in the choice of proposed analysis.

The difference between lunar laser and interplanetary radio range is multifold. A typical uncertainty of recent lunar laser range data is around $1\,{\rm cm}$, which is about two orders of magnitude better than that of current interplanetary radio range measurements~\cite{2021AJ....161..105P}. However, the peak sensitivity in both studies appears around $m=10^{-15}\eV$ at which the dark matter wavelength is $\lambda_{\rm DM} = 1/(m\sigma) \sim 2\,{\rm AU}$. While a typical interplanetary range is around the same order as the wavelength, a nominal lunar range is deep inside the tidal regime, where the signal is suppressed by an additional factor of $L / \lambda_{\rm DM} \sim 10^{-3}$. 

The equations used to study the evolution of the test masses are also different. In this work, we use the non-relativistic limit of the geodesic equations of two test masses in the coordinate system defined with the conformal Newtonian gauge. In Ref.~\cite{Foster:2025csl}, the equation used for the test mass evolution appears to be the one given in the Fermi normal coordinate of the center of mass frame, $d^2 L^i / d\tau^2= -  R_{0i0j}  L^j = - ( \ddot\Psi  \delta_{ij} +\partial_i \partial_j \Phi)  L^j $, which is an approximation in the tidal limit.\footnote{In Ref.~\cite{Foster:2025csl}, only $\ddot\Psi$ is considered. This is expected to be a good approximation for the coherent density fluctuation. For stochastic fluctuations, the other contribution $\partial_i \partial_j \Phi$ is expected to provide a larger signal in range measurements.} Although the form of the equation is different, a relativistically defined two-way range observable---proper time lapse between the transmission and reception event of the range signal---is the same in both coordinate systems in the tidal limit. It turns out that the two-way range observable  in the Fermi normal coordinate is simply given by $L^i$. 

In the binary system, the two-way range observable can be expressed in terms of osculating orbital elements since it is simply given by the binary separation. This allowed authors of Ref.~\cite{Foster:2025csl} to solve the equation of osculating orbital elements perturbatively to compute the signal of ultralight dark matter for a time period longer than a typical orbital time period. For a multi-planet system, the range will be a complicated function of the orbital parameters of different planets. For this reason, we directly solve the Newtonian equation for test masses and compute the signal numerically and analytically by limiting the duration of each time series to be smaller than a year. 

We may also compare our results with those obtained by using pulsar timing and astrometry. The peak sensitivity of the range measurement appears at the mass range, which is two orders of magnitude larger than the location of the peak sensitivity in pulsar timing and astrometry. This is due to the segmentation of the time series that we assume in the analysis. Even though the total time baseline of the interplanetary range is comparable to or longer than that of pulsar timing and astrometry, dividing the time series into a $6$-month-long segment limits the lowest possible mass that can be probed by the proposed analysis. In particular, since we assume that at least a few stochastic fluctuations are sampled in the data stream, the lowest possible mass that can be probed in this way will be given by $m \sim \sigma^2 / T_{\rm seg}$. An interesting open question is whether we can extend the sensitivity beyond this limit, for instance, by generalizing the approach adopted in Ref.~\cite{Foster:2025csl} to interplanetary range measurements to search for signals of duration longer than a typical orbital period of the planets. We do not address this question in this work and leave it as future work.

\section{Conclusion}\label{sec:Conclusion}
This work attempts to address the question of whether one can use precision solar system range measurements for an ultralight dark matter search. Given the lack of observational evidence of dark matter on a scale of the solar system, the current best limit on ultralight dark matter density is achieved via the analysis of pulsar timing measurements~\cite{Kim:2023kyy}. As the sensitivity of timing measurement hinges on the precise modeling of the solar system barycenter, it is natural to ask if the underlying measurements for constructing planetary ephemerides could be used for a similar ultralight dark matter search. 

To address this question, we first identified range measurements as observables for this study. The distances between Earth and various planets, e.g. Mercury, Venus, Mars, Saturn, and Jupiter, are typically measured to the ${\cal O}(1)\,{\rm m}$ level of uncertainty. By characterizing the ULDM-induced signal in the auto- and cross-correlation of range measurements, we derived a projected sensitivity by assuming typical noise characteristics of such range measurements. With the benchmarks we have chosen, we found that one can probe $\bar \rho \sim 10^5 \rho_0$ at $m = 10^{-15} \eV$, where the peak sensitivity occurs. 

Due to the non-linearity of the gravitating system, it is nontrivial to characterize the signal for intervals beyond typical orbital time scales. For this reason, we proposed an analysis that uses a segmented time series of range measurements. This resembles the analysis of stochastic gravitational wave search in the ground-based interferometers~\cite{LIGOScientific:2003jxj, LIGOScientific:2006zmq, LIGOScientific:2019vic, KAGRA:2021kbb, LIGOScientific:2025bgj}. By segmenting the time series into smaller pieces with a time span shorter than a few months, the planetary system can be approximated in a way that the gravitational force due to the Sun and other planets can be modeled as a linear and quadratic drifts. The theoretical auto- and cross-spectral densities are obtained under this assumption, and confirmed by a dedicated numerical simulation of the solar system. The spectrum, as well as the correlation coefficients, exhibits a reasonable agreement between analytic and numerical computation.

This work can be extended in several directions. While we have limited the analysis by only considering range measurements, there are astrometry and very-long baseline interferometry measurements that provide complementary information on planetary trajectories. It would be interesting to examine how these measurements can be included and how they impact the projected sensitivity on ultralight dark matter density. In addition, we derived the projected sensitivity based on the cross-correlation statistic with a white noise spectrum that reflects the noise level of current range measurements. An obvious further direction for exploration is to analyze actual residual range measurements for the ULDM search, which we leave for future work. 

\begin{acknowledgments}
This work was supported by the Deutsche Forschungsgemeinschaft under Germany’s Excellence Strategy - EXC 2121 Quantum Universe - 390833306.
JF is supported by an ERC StG grant (“AstroDarkLS”, Grant No. 101117510).
Some numerical computations have been performed at the Vera cluster supported by the Italian Ministry of Research and by Sapienza University of Rome.
\end{acknowledgments}

\appendix

\section{Signal}\label{app:signal}
We provide detailed derivations on ultralight dark matter signals in two-range measurements and their power spectral densities. In Appendix~\ref{app:signal_derivation}, we first discuss how the dark matter signal enters the two-way range measurement, and in Appendix~\ref{app:Computation_Spectrum}, we compute the power spectra for the corresponding dark matter signal.

\subsection{Signal Derivation}\label{app:signal_derivation}
\begin{figure}[!t]
\centering
\includegraphics[width=0.4\textwidth]{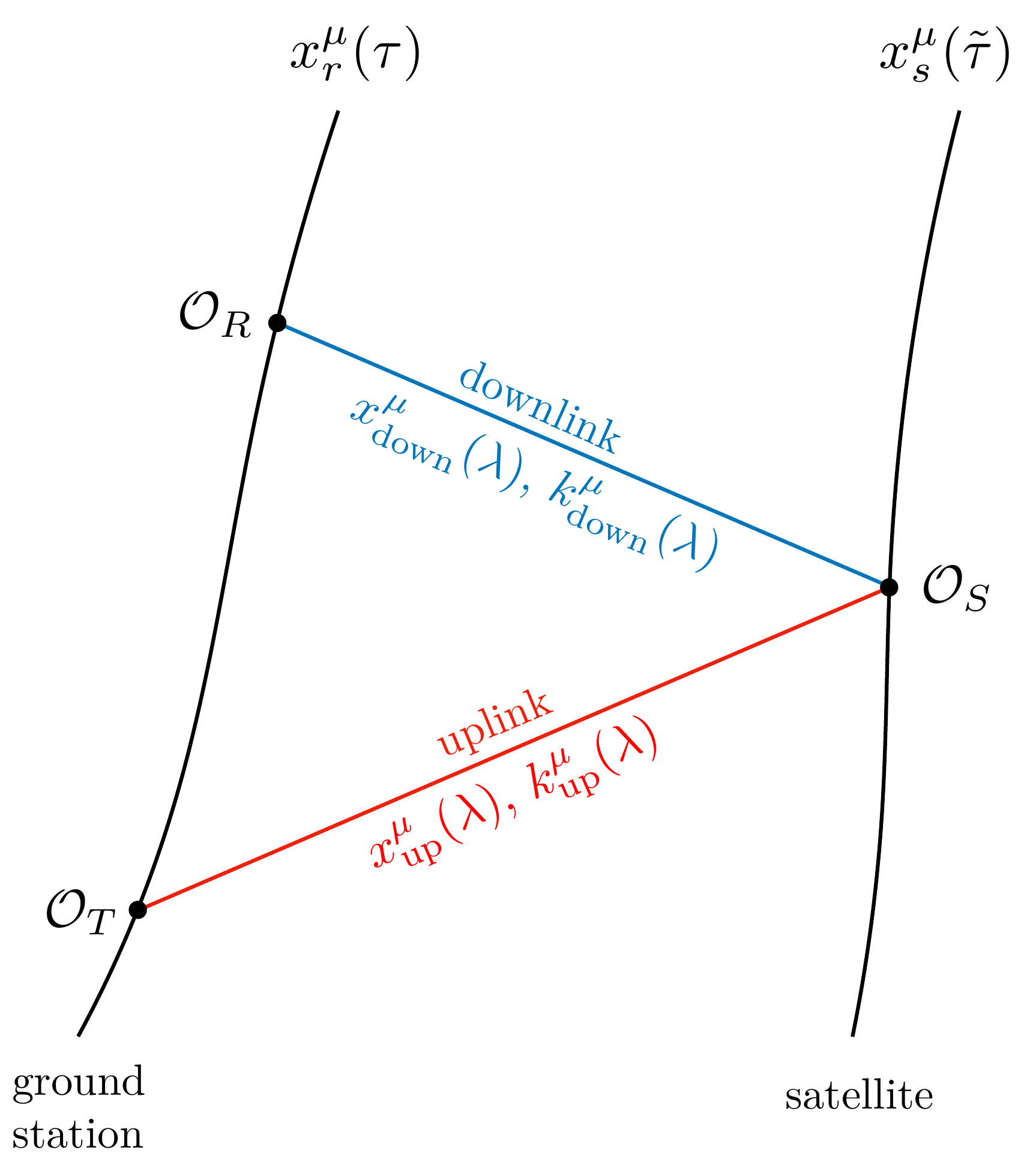}
\caption{Schematic figure describing the worldlines of the ground station and the satellite, as well as the photon. The radio wave is transmitted from the ground station at the proper time $\tau_t$, which corresponds to the coordinate time $t_t$ and the photon affine parameter $\lambda_t$. It reaches the satellite at $t_s$, at which the transponder sends back the signal to the ground station. The downlink signal is received at the same ground station at $t$.}
\label{fig:worldlines}
\end{figure}

Consider the following setup: a radio wave is transmitted at the ground station at its proper time $\tau_t$. This proper time corresponds to coordinate time $t_t$, and the affine parameter of the photon $\lambda_t$. The radio wave reaches the satellite at $t_s$, and the transponder sends the signal back to the ground station. This downlink signal is received by the same ground station at its proper time $\tau$. From this measurement, the range can be estimated as 
$$
L(\tau) = \frac{\tau - \tau_t}{2} = \frac{\Delta \tau(\tau) }{2}\;,
$$ 
where $\Delta\tau(\tau)$ is the proper time lapse between two events: the transmission ${\cal O}_T$ and the reception  ${\cal O}_R$ of the range signal. A schematic diagram of the setup is provided in Fig.~\ref{fig:worldlines}.

The round-trip light travel time can be written in terms of coordinate time, 
$$
\Delta \tau =
\Delta t 
+ \int_{t_t}^t dt' 
\bigg(
    \Phi
    - \frac{1}{2} v^2 
\bigg) \;, 
$$
where $v$ is the velocity of the observer at the ground station. The integral is often referred to as the Einstein delay. It includes an ultralight dark matter signal. The integral should be carried out along the worldline of the observer.

The dark matter signals also reside in the coordinate time lapse. The coordinate time lapse during the round-trip can be written as
\begin{align}
\Delta t= (t - t_s) + (t_s - t_t)\;. 
\end{align}
The computation of each time lapse can be obtained by integrating the geodesic equation of the photon perturbatively. Consider the uplink path. The computation for the downlink path can be done in the same way. We first expand the photon four-momentum as 
$$
k_{\rm up}^\mu = k^\mu_{{\rm up},0} + \delta k_{\rm up}^\mu\;,
$$ 
where $k_{{\rm up},0}^\mu = \omega_\gamma ( 1 , \hat{\boldsymbol n})^\mu$ is a photon four-momentum with the photon frequency $\omega_\gamma$ and the unit vector $\hat{\boldsymbol n}$, and $\delta k^\mu_{\rm up}$ is a perturbed four-momentum that is of linear order in the metric perturbation. By integrating the geodesic equation, we find
\begin{align}
\!\! k_{\rm up}^\mu(\lambda_s) &= k_{\rm up}^\mu(\lambda_t) 
- \int^{\lambda_s}_{\lambda_t} d\lambda \, \Gamma^\mu_{\;\nu\rho} k^\nu_{{\rm up},0} k^\rho_{{\rm up},0}  \;, 
\\
\!\! x_{\rm up}^\mu(\lambda_s) &= x_{\rm up}^\mu(\lambda_t) + k_{\rm up}^\mu(\lambda_t)(\lambda_s - \lambda_t) \nonumber\\
&\qquad\qquad 
- \int^{\lambda_s}_{\lambda_t} d\lambda' \int^{\lambda'}_{\lambda_t} d\lambda'' \, \Gamma^\mu_{\;\nu\rho} k^\nu_{{\rm up},0} k^\rho_{{\rm up},0} \;. 
\label{uplink_x}
\end{align}
The boundary conditions are
\begin{align}
&
{\cal O}_T : \, x_r^{\mu}(\tau_t) = x^\mu_{\rm up}(\lambda_t) \;,
\label{bT}
\\
&
{\cal O}_S : \, x_s^{\mu}(\tilde \tau_s) = x^\mu_{\rm up}(\lambda_s) = x^\mu_{\rm down}(\lambda_s) \;, 
\label{bS}
\\
&
{\cal O}_R : \, x_r^\mu(\tau) = x^\mu_{\rm down}(\lambda) \;,
\label{bR}
\end{align}
i.e. the worldlines of the range photon and ground station must intersect at the transmission ${\cal O}_T$ and reception ${\cal O}_R$, and likewise for the worldlines of the range photon and the satellite at the satellite arrival event ${\cal O}_S$. 

The coordinate time lapse $(t_s-t_t)$ can be obtained from the expression of the uplink photon path $x^\mu_{\rm up}(\lambda)$. Taking the inner product between the equation \eqref{uplink_x} and $-\eta_{\mu\nu} k_{{\rm up},0}^\nu / \omega_\gamma$, we find
\begin{align}
\!\!\!
t_s - t_t 
=& \, \hat{\boldsymbol n} \cdot \big[ \boldsymbol x_s(t_s) - \boldsymbol x_r(t_t) \big]
\nonumber\\
\!\!\!
& + \frac{1}{2\omega_\gamma} \int^{\lambda_s}_{\lambda_t} d\lambda \, \delta g_{\mu\nu}(\lambda) k_{{\rm up},0}^\mu k_{{\rm up},0}^\nu 
\nonumber\\
\!\!\!
& - \frac{(\lambda_s - \lambda_t)}{2\omega_\gamma} ( 2\eta_{\mu\nu} k^\mu_{{\rm up}, 0} \delta k_{\rm up}^\nu + \delta g_{\mu\nu} k_{{\rm up}, 0}^\mu k^\nu_{{\rm up}, 0} )|_{\lambda_t} 
\nonumber\\
\!\!\!
=& \,
\hat{\boldsymbol n} \cdot \big[ \boldsymbol x_s(t_s) - \boldsymbol x_r(t_t) \big]
- 2 \omega_\gamma \int^{\lambda_s}_{\lambda_t} d\lambda' \, \Phi(x_{\rm up}(\lambda'))\;. 
\end{align}
The last term in the first equality vanishes due to the null condition on the four-momentum, $0=g_{\mu\nu} k^\mu k^\nu= 2 \eta_{\mu\nu} k^\mu_0 \delta k^\nu + \delta g_{\mu\nu} k^\mu_0 k^\nu_0$. We also used the boundary conditions \eqref{bT}--\eqref{bS}. Repetition of the same computation for the downlink signal results in a similar expression for $t-t_s$. Combining all terms, we find
\begin{align}
\Delta L(\tau) = \Delta L^{O}(\tau) + \Delta L^S(\tau) + \Delta L^E(\tau)\;,
\end{align}
where each term represents the dark matter induced fluctuations in the two-way range measurements:
\begin{align}
\!\! 
\Delta L^O 
&= 
\frac{1}{2} \hat{\boldsymbol n}
\cdot
\big[
    (\boldsymbol x_s(t_s) - \boldsymbol x_r(t_t))
    + (\boldsymbol x_s(t_s) - \boldsymbol x_r(t))
\big] \;, 
\\
\!\!
\Delta L^S 
&= 
- \omega_\gamma
\bigg[
    \int_{\lambda_s}^{\lambda} d\lambda' 
    + \int_{\lambda_t}^{\lambda_s} d\lambda' 
\bigg] \Phi(\lambda') \;, 
\\
\!\!
\Delta L^E 
&= \frac{1}{2}
\int^t_{t_t} dt' \Big( \Phi - \frac{1}{2} v^2 \Big)\; . 
\end{align}
This corresponds to Eqs.~\eqref{tau_O}--\eqref{tau_E} in the main text. Each effect will be quantified in the following subsection by its respective power spectrum.

\subsection{Signal Spectrum}\label{app:Computation_Spectrum}
In this subsection, we present detailed computations of the signal power spectrum. The goal of this Appendix is to characterize the signal in terms of a one-sided power spectrum, defined as
\begin{align}
\big\langle 
    \widetilde{\Delta L}_a(f) 
    \widetilde{\Delta L}_b^*(f')
\big\rangle
= \frac{1}{2} \delta(f-f') \Sigma_{ab}(f)\;,
\end{align}
where the subscript $a$ indexes different two-way range measurements. We will compute the signal power spectrum due to (i) orbital perturbation, (ii) Shapiro delay, and (iii) Einstein delay. 

We introduce a few useful quantities with which we express our signal power spectrum before we proceed to compute the signal power spectrum. They are (i) the frequency spectrum of a single test mass, (ii) and of the potential $\Phi$ itself. They are defined as
\begin{align}
\langle 
    \widetilde\Phi(f, \boldsymbol x)
    \widetilde\Phi^*(f', \boldsymbol x)
\rangle
= \frac{1}{2} \delta(f-f') S_\Phi(f)\;,
\\
\langle 
    \widetilde{\delta x}_n(f, \boldsymbol x)
    \widetilde{\delta x}_n^*(f', \boldsymbol x)
\rangle
= \frac{1}{2} \delta(f-f') S_x(f)\;,
\end{align}
where $\delta x_n = \boldsymbol{\delta x} \cdot \hat {\boldsymbol n}$ is a fluctuation in the position of the test mass projected onto an arbitrary direction $\hat{\boldsymbol n}$. The frequency spectrum of the potential can be directly computed from its four-dimensional power spectrum:
\begin{align}
S_\Phi(f) 
&= 2 \int \frac{d^3k}{(2\pi)^3} P_\Phi(k) 
= \frac{(4\pi G \bar \rho)^2}{m^5 \sigma^6} \frac{K_1(\omega \tau)}{\omega\tau}\;.
\label{S_Phi}
\end{align}
Here, we use the angular frequency $\omega = 2 \pi f$ on the right-hand side for notational brevity. The potential power spectrum is related to that of the density contrast in Eq.~\eqref{eq:dens_spec} via $P_\Phi(k) = (4\pi G / k^2)^2 P_{\delta\rho}(k)$. 
For a single test particle that is subject to the perturbation due to ultralight dark matter, the equation of motion is given by 
$$
    \ddot{\boldsymbol{\delta x}} = - \nabla \Phi\;. 
$$
The Fourier expression becomes
\begin{align}
\widetilde{\delta x}_n(f)
= 
\int \frac{d^3k}{(2\pi)^3}
e^{i \boldsymbol k \cdot \boldsymbol x}
\frac{i \boldsymbol k \cdot \hat{\boldsymbol n}}{\omega^2}
\widetilde{\Phi}(k)  \;. 
\end{align}
The resulting one-sided power spectrum is
\begin{align}
S_x(f) 
&= 
\frac{2}{\omega^4} 
\int \frac{d^3k}{(2\pi)^3} k^2 P_\Phi(k) 
(\hat{\boldsymbol k} \cdot \hat{\boldsymbol n})
(\hat{\boldsymbol k} \cdot \hat{\boldsymbol n})
\nonumber\\
&=
\frac{2}{3\omega^4} 
\frac{(4\pi G \bar \rho)^2}{m^3 \sigma^4} K_0(\omega \tau)\;.
\label{sigma_deltax}
\end{align}
For both expressions, we have used the power spectrum $P_\Phi(k)$ obtained with an isotropic dark matter velocity distribution. The signal power spectrum in the following subsections will be expressed in these quantities. 

These expressions can be understood within a quasiparticle picture of ultralight dark matter. The root-mean-square fluctuation of the position and the potential averaged over $T$ are given by
\begin{align}
\delta x_{n, \rm rms} 
&= 
\bigg[ \int_{f_\ell}^{f_u} df S_x(f) \bigg]^{\frac12}
\approx
a_q \tau^2 
\left[
\frac{2}{9\pi^2} \frac{T^3}{\tau^3} \ln\frac{T}{\pi\tau} 
\right]^{\frac12}\;,
\\
\Phi_{\rm rms} 
&= 
\bigg[ \int_{f_\ell}^{f_u} df S_{\Phi}(f) \bigg]^{\frac12}
\approx
2 \Phi_q \Big(\frac{T}{\tau}\Big)^{\frac12}\;,
\end{align}
where $f_\ell = 1/T$, $f_u =1/2\Delta t$, $T$ is the integration time, and $\Delta t$ is the cadence of observation. We introduce the acceleration and the potential due to a quasiparticle, $a_q = GM_{\rm eff}/\lambda_\text{DM}^2$ and $\Phi_q = GM_{\rm eff}/\lambda_\text{DM}$. The second expression in each line is obtained in the limit $T \gg \tau$. Recall that $M_{\rm eff} = \bar \rho \lambda_\text{DM}^3$ and $\lambda_\text{DM} = 1/ m \sigma$; the rms fluctuation of a test mass position and the potential over a coherence time scale can be simply estimated by treating the dark matter density fluctuation as a quasiparticle of its mass $M_{\rm eff}$ and size $\lambda_\text{DM}$. 

\subsubsection{Orbital Perturbation Signal}\label{app:signal_calc}
The round-trip light travel time fluctuates as ultralight dark matter perturbs Earth and each satellite. The effect is given by Eq.~\eqref{tau_O}, i.e.,
\begin{align}
\Delta L_a(t) =& 
\frac12
\Big(
\hat{\boldsymbol n}_a \cdot 
\big[ 
    \boldsymbol{\delta x}_a(t - \bar L_a) 
    - \boldsymbol {\delta x}_r(t)
\big]
\nonumber\\
&
+ \hat{\boldsymbol n}_a \cdot 
\big[ 
    \boldsymbol{\delta x}_a(t - \bar L_a) 
    - \boldsymbol{\delta x}_r(t-2 \bar L_a) 
\big]
\Big)\;,
\nonumber
\end{align}
where $\bar L_a$ is the nominal separation, and $\hat{\boldsymbol n}_a$ is a unit vector from the ground station to the satellite $a$.

The computation of the signal power spectrum is essentially the same as the computation of the frequency spectra of the test mass position and the potential. The starting point is to find a Fourier expression of $\Delta L_a(t)$. Using Eq.~\eqref{tau_O} and the equation of motion for the test mass, we find
\begin{align}
\widetilde{\Delta L_a}(f)
= \int \frac{d^3k}{(2\pi)^3}
e^{i \boldsymbol k \cdot \boldsymbol x_r}
\frac{i {\boldsymbol k} \cdot \hat{\boldsymbol n}_a}{\omega^2}
D_a(f, \boldsymbol k)
\widetilde\Phi(f,\boldsymbol k)\;,
\end{align}
where $\boldsymbol x_r$ is the nominal position of the reference observer, and $D_a(f,\boldsymbol k)$ is the response function, defined as
\begin{align}
D_a(f,\boldsymbol k)
=&
\frac{1}{2}
\bigg[ 
\big(  e^{i (\omega \bar L_a + \boldsymbol k \cdot \bar{\boldsymbol L}_a)} - 1\big)
\nonumber\\
&
+ \big( e^{i (\omega \bar L_a + \boldsymbol k \cdot \bar{\boldsymbol L}_a)} -  e^{2i\omega \bar L_a} \big)
\bigg] \;. 
\end{align}
Each term in the round parentheses represents the response of the downlink and uplink signal with respect to dark matter fluctuation. The above Fourier representation of the round-trip light travel time allows us to write the power spectrum in terms of the response function and the power spectrum of the potential:
\begin{align}
\Sigma_{ab}(f)
= \frac{2}{\omega^4} 
\int \frac{d^3k}{(2\pi)^3}
k^2 P_\Phi(k) 
(\hat{\boldsymbol k} \cdot \hat{\boldsymbol n}_a)
(\hat{\boldsymbol k} \cdot \hat{\boldsymbol n}_b)
D_a D_b^*
\;. 
\end{align}
The arguments of the response function $D_{a,b}(f,\boldsymbol k)$ are suppressed for brevity.

The spectrum can be computed explicitly. With Eq.~\eqref{eq:dens_spec}, a straightforward computation reveals the following:
\begin{align}
\Sigma_{ab}(f) 
&= 
S_{x}(f) {\cal I}_{ab}(f, \bar{\boldsymbol L}_a, \bar{\boldsymbol L}_b)\;,
\end{align}
where $S_x(f)$ is the power spectrum for the single test mass fluctuation from Eq.~\eqref{sigma_deltax}, and the response integral is defined as
\begin{align}
{\cal I}_{ab}
= 3 \int_0^\infty dx \, p_1(x) 
\int \frac{d\Omega}{4\pi} 
(\hat{\boldsymbol k} \cdot \hat{\boldsymbol n}_a) 
(\hat{\boldsymbol k} \cdot \hat{\boldsymbol n}_b) 
D_a D_b^* \;,
\label{eq:resp_int}
\end{align}
where 
\begin{align}\label{eq:pn}
p_n(x) = 
\frac{ x^{-n} e^{- \frac{x^2}{4} - \frac{(\omega\tau)^2}{x^2} } }{\int_0^\infty dx \, x^{-n} e^{ - \frac{x^2}{4} - \frac{(\omega\tau)^2}{x^2}} }\;. 
\end{align}
We have changed the momentum integration over $k$ into the dimensionless variable $x=|\boldsymbol{k}|/(m\sigma)$. The dimensionless  integral encodes the response of the two-way range measurement with respect to ULDM-induced orbital perturbations. The angular part of the integral can be performed explicitly:
\begin{widetext}
\begin{align}\label{eq:angular}
\int \frac{d\Omega}{4\pi} 
(\hat{\boldsymbol k} \cdot \hat{\boldsymbol n}_a) 
(\hat{\boldsymbol k} \cdot \hat{\boldsymbol n}_b) 
D_a D_b^* 
= &
e^{i \omega(\bar L_a - \bar L_b)}
\Bigg[
(\hat{\boldsymbol n}_a \cdot \hat{\boldsymbol n}_b)
\bigg(
    \frac{\cos(\omega \bar L_a)\cos(\omega \bar L_b)}{3} 
    - \cos(\omega \bar L_b) \bigg( \frac{j_1(\rho_a)}{\rho_a} - j_2(\rho_a) \bigg)
\nonumber\\    
&
    - \cos(\omega \bar L_a) \bigg( \frac{j_1(\rho_b)}{\rho_b} - j_2(\rho_b) \bigg)
    + \bigg( \frac{j_1(\rho_{ab})}{\rho_{ab}} - j_2(\rho_{ab}) \bigg)
\bigg)
+  \frac{\bar L_a \bar L_b}{\bar L_{ab}^2} 
( 1 -(\hat{\boldsymbol n}_a \cdot \hat{\boldsymbol n}_b)^2 ) j_2(\rho_{ab})
\Bigg]\;,
\end{align}
\end{widetext}
where $\rho_{a,b}= |\boldsymbol{k}| \bar L_{a,b}$, $\rho_{ab} = |\boldsymbol{k}| \bar L_{ab}$, $\bar L_{ab} = |\bar{\boldsymbol L}_a - \bar{\boldsymbol L}_b|$, and $j_n(x)$ is a spherical Bessel function of the first kind. 

It is instructive to examine a few limits of the above expressions. First, note that $\omega \bar L \ll 1$ in our case; that is, ultralight dark matter does not fluctuate significantly at a few minutes scale $t \sim {\cal O}(1)\,{\rm AU} \sim {\cal O}(1) \,{\rm min}$. In addition, let us only consider the auto-correlation $a=b$. In the short (long) wavelength limit $m \sigma \bar L_a \gg 1$ ($m \sigma \bar L_a \ll 1$), we find
\begin{align}
\frac{\Sigma_{aa}(f)}{S_x(f)}
\approx 
\begin{cases}
(m \sigma \bar L_a)^2
\frac{6 \omega \tau  K_1(\omega \tau)}{5K_0(\omega\tau)}
&
m \sigma \bar L_a \ll 1
\\
2-\frac{\Gamma(0,\omega^2 \bar L_a^2/\sigma^2)}{K_0(\omega\tau)}
&
m \sigma \bar L_a \gg 1
\end{cases} \;,
\label{limit_Orb}
\end{align}
where $\Gamma(a,b)$ is an incomplete Gamma function. The expression for the long wavelength limit includes an additional suppression factor of $(\bar L_a/ \lambda_\text{DM})^2$, which manifests the tidal suppression of the differential acceleration. The expression for the short wavelength limit is approximately twice the power spectrum of a single test mass position for sufficiently high frequencies $\omega > \sigma/\bar L_a$. For $\omega < \sigma/\bar L_a$, we obtain $\mathcal{O}(1)$ deviations from 2; in the quasi-particle picture, this can be seen by observing that a fluctuation takes a time $\delta t \sim \bar L_a/\sigma$ to propagate between the two test masses and thus, it will affect the frequency spectrum below $\omega \sim 1/\delta t=\sigma/\bar L$.

Let us investigate the origin of this behavior more rigorously. In the limit $\omega \bar L_a\ll 1$, the angular integral of the response function Eq.~\eqref{eq:angular} simplifies to
\begin{align}
\Bigg[
    \frac{2}{3} 
    - 2 \bigg( \frac{j_1(\rho_a)}{\rho_a} - j_2(\rho_a) \bigg)
\Bigg]\;.
\end{align}
The first term can be integrated over $x$ [cf.\ Eq.~\eqref{eq:resp_int}], giving rise to $2$ in the short wavelength limit as obtained in Eq.~\eqref{limit_Orb}. On the other hand, the second term in the round parentheses takes a non-vanishing value only when the argument $\rho_a = k \bar L_a = x (m \sigma \bar L_a) \ll 1$. The correction to the response integral due to this term is therefore
\begin{align}
\Delta {\cal I}_{ab} 
&\approx
-\frac{2}{K_0(\omega \tau)}
\int_0^{1 / (m\sigma \bar L_a)}  \frac{dx}{x} 
\exp\left[ - \frac{(\omega \tau)^2}{x^2} \right]
\nonumber \\
&=
-\frac{\Gamma(0,\omega^2 \bar L_a^2 /\sigma^2)}{K_0(\omega \tau)}\;,
\end{align}
where we have already approximated $j_1(\rho_a)/\rho_a - j_2(\rho_a) \approx 1/3$ in the first line. 

The physical interpretation of this correction is clear. We assume that the momentum of dark matter follows a normal distribution. It is always possible to find momentum modes $k < 1 / \bar L_a$ that could still introduce a correlation in the perturbation of the reference and the test mass $a$, correcting the short wavelength limit of the response integral. Furthermore, due to the long-range nature of the gravitational interaction, each decade of $k$ seems to contribute to the integral equally as long as $k > \omega / \sigma$. For modes with even smaller wavenumber, $k < \omega / \sigma$, the integral is Boltzmann suppressed; given that the stochastic density fluctuation is a result of interference of two plane waves, $\boldsymbol k = \boldsymbol k_1 - \boldsymbol k_2$ and $\omega = \omega_1 - \omega_2 = (k_1 + k_2) k / 2m$, and thus, a small wavenumber $k < \omega / \sigma$ means $k_1 + k_2 > m \sigma$, which is Boltzmann suppressed. 

We may estimate the rms fluctuations of the round-trip light travel time. In the short wavelength limit, the power spectrum is identical to that of a single test mass fluctuation up to a constant. Therefore, the rms fluctuation over a coherence time scale can be estimated as in the previous section:
$$
(\Delta L_a)_{\rm rms}
= \bigg[ \int df \, \Sigma_{aa}(f) \bigg]^{\frac12} \sim a_q \tau^2 \;. 
$$
where $a_q = GM_{\rm eff}/ \lambda^2$ is the acceleration due to a quasiparticle. This justifies the estimation in the main text for the dark matter signal from the orbital perturbation, based on the quasiparticle heuristic.

\subsubsection{Shapiro Delay Signal}
The Shapiro delay signal is given by Eq.~\eqref{tau_S}:
$$
\Delta L_a(t)
= - \omega_\gamma
\bigg[
\int_{\lambda_s}^{\lambda} d\lambda' 
+ \int_{\lambda_t}^{\lambda_s} d\lambda'
\bigg] \Phi(x^\mu(\lambda))\;,
$$
where the argument of the potential is the nominal photon worldline. The photon four-momentum along the uplink and downlink is $k_0^\mu = \omega_\gamma(1, \pm \hat{\boldsymbol n}_a)$, and thus, the variation of $x^\mu$ is simply given by $\Delta x^\mu(\lambda) = k_0^\mu \Delta \lambda  =\bar L_a(1, \pm \hat{\boldsymbol n}_a)$. By explicitly performing the $\lambda$-integral, we find
\begin{align}
\widetilde{\Delta L}_a(f)
= \bar L_a \int \frac{d^3k}{(2\pi)^3} 
e^{i \boldsymbol k \cdot \boldsymbol x_r} 
D_a(f,\boldsymbol k) \widetilde \Phi(k)\;. 
\end{align}
where the dimensionless response function is given by
\begin{align}
D_a(f,\boldsymbol k) 
= &
i
\left[
\frac{e^{i (\omega \bar L_a + \boldsymbol k \cdot \bar {\boldsymbol L}_a)} - 1}{\omega \bar L_a + \boldsymbol k \cdot \bar {\boldsymbol L}_a}
+ 
\frac{e^{2i\omega \bar L_a} - e^{i (\omega \bar L_a + \boldsymbol k \cdot \bar{\boldsymbol L}_a)}}{\omega \bar L_a - \boldsymbol k \cdot \bar{\boldsymbol L}_a}
\right] . 
\end{align}
With the above form of the Fourier component and the response function, the cross-spectral density can be written as
\begin{align}
\Sigma_{ab}(f)
= 2 \bar L_a \bar L_b \int \frac{d^3k}{(2\pi)^3} P_\Phi(k) D_a D_b^*  .
\end{align}
By following the same procedure as in the computation of the orbital perturbation signal power spectrum, we find
\begin{align}
\Sigma_{ab}(f)
&=  S_\Phi(f) \bar L_a \bar L_b {\cal I}_{ab}(f, \bar {\boldsymbol L}_a, \bar {\boldsymbol L}_b)  ,
\end{align}
where $S_\Phi(f)$ is given in Eq.~\eqref{S_Phi}, and the dimensionless response integral ${\cal I}_{ab}(f,\bar{\boldsymbol L}_a, \bar{\boldsymbol L}_b)$ are
\begin{align}
{\cal I}_{ab}(f)
&= \int_0^\infty dx \, p_3(x)
\int \frac{d \Omega}{4\pi} D_a D_b^* , 
\end{align}
where the arguments of ${\cal I}_{ab}$ and its integrand are suppressed for brevity. 

It is helpful to take a few limits to examine the behavior of the frequency spectrum. Consider the response integral for the auto-correlation $a=b$. In the long- and short-wavelength limit, we find
\begin{align}
\frac{\Sigma_{aa}(f)}{S_\Phi(f) \bar L_a^2}
\approx  
\begin{cases}
4
& m \sigma \bar L_a \ll 1 
\\
\frac{2 \pi^{3/2} e^{-\omega\tau}(1 + \omega\tau)}{(\omega\tau)^2 K_1(\omega\tau)} \frac{1}{m\sigma \bar L_a} 
&  m \sigma \bar L_a \gg 1 
\end{cases} . 
\label{shapiro_response_approx}
\end{align}
To derive this expression, we use $\omega \bar L_a \ll 1$ as before. Then the square of the response function can be approximated as $|D_a|^2 \approx 4 \,{\rm sinc}^2\,(\boldsymbol k \cdot \bar{\boldsymbol L}_a/2)$ where ${\rm sinc}\,(x) = \sin x /x$. Using this, the response integral can be obtained as Eq.~\eqref{shapiro_response_approx} in each limit. 

The interpretation of the above result is straightforward. The root mean square fluctuation of the round-trip light travel time is
\begin{align}
\Delta L_{a, \rm rms}
&= 
\bigg[ \int_{f_\ell}^{f_u} df' \, \Sigma_{aa}(f)\bigg]^{1/2}
\nonumber\\
&\sim 2 \Phi_{\rm rms} 
\begin{cases}
\bar L_a
& m \sigma \bar L_a \ll 1
\\
\lambda ( \bar L_a / \lambda_{\rm DM})^{1/2}
& m \sigma \bar L_a \gg 1
\end{cases}
\end{align}
where $\Phi_{\rm rms}^2 = \int_{f_\ell}^{f_u} df \, S_\Phi(f)$ and the second expression is obtained for $T \sim {\cal O}(1)\tau$. In the long wavelength limit, the Shapiro time delay coherently adds up as the potential does not change significantly during the round-trip, while in the short wavelength limit, it only incoherently adds up, which is manifest via the $(\bar L_a/\lambda_{\rm DM})^{1/2}$ factor. This estimation also justifies the approximation given in the main text.

\subsubsection{Einstein Delay Signal}
The Einstein delay signal is given by Eq.~\eqref{tau_E}:
$$
\Delta L_a(t)
= \frac{1}{2} \int^t_{t_t} dt' \Big( \Phi - \frac{v^2}{2} \Big) . 
$$
For the moment, we ignore the second term. Although the observer velocity is also affected by dark matter, $\boldsymbol v = \boldsymbol v_0 + \boldsymbol{\delta v}$, the velocity perturbation induced by dark matter is multiplied by the observer velocity, and therefore is small compared to the first term. We will justify this statement shortly. 

The above integral can be trivially done. The time lapse between the transmission and reception of the range signal is typically on the scale of a few minutes, $t - t_t \sim {\cal O}(1)\,{\rm min}$, while the potential oscillates at 
$$
\omega 
\sim m \sigma^2 
= 4\times 10^{-7}\,{\rm Hz} \times \Big( \frac{m}{10^{-15}\eV} \Big).
$$
As the potential remains almost constant over $t-t_t$, we may approximate
$$
\Delta L_a(t) \approx \frac{1}{2} \bar L_a \Phi(t, \boldsymbol x_r) ,
$$
from which the cross-spectral density is computed as
\begin{align}
\Sigma_{ab}(f)
&\approx \frac{1}{4} S_\Phi(f) \bar L_a \bar L_b  . 
\end{align}
The result is identical to the Shapiro delay signal of the long wavelength limit upto a constant numerical factor. 

In the previous discussion, we ignored a potential effect from the velocity perturbation of the observer induced by ultralight dark matter:
\begin{align}
\Delta L_a(t) = 
- \frac{1}{2} \int^t_{t_t} dt' \, \boldsymbol v_0 \cdot \boldsymbol{\delta v}(t').
\end{align}
This approximation can be justified as follows. Solving the geodesic equation perturbatively, one finds that the perturbation can be written as
\begin{align}
\boldsymbol{\delta v}(t) \approx - \int^t_{t_t} dt' \, \nabla \Phi 
\sim {\cal O}(m \sigma\bar L) \Phi
\end{align}
and thus, as long as $v_0 (m \sigma \bar L) < 1$, this effect of the observer velocity perturbation is subdominant compared to that of the potential.

\subsubsection{Comparison of Each Signal}
Let us carefully compare each signal power spectrum. Since the Einstein time delay spectrum is identical to that of the Shapiro time delay in the long wavelength limit, we only consider the signal spectrum of the orbital perturbation and the Shapiro time delay. In addition, we will focus on frequencies smaller than the inverse of the coherence time, $\omega < 1 /\tau$, as we expect that most of the signal will arise in this region. We find in each limit
\begin{align}
\frac{\Sigma^{\rm (S)}_{aa}(f)}{\Sigma^{\rm (O)}_{aa}(f)}
\approx 
\sigma^4
\begin{cases}
5 ( \omega \tau )^2
&
m \sigma \bar L_a \ll 1
\\
\frac{3 \pi^{3/2}}{2 \ln(1/\omega\tau)} (\omega \tau) ( m \sigma \bar{L}_a)
&
m \sigma \bar L_a \gg 1
\end{cases},
\end{align}
where the superscripts (S) and (O) denote the Shapiro time delay signal and the signal from orbital perturbation, respectively. Both expressions are obtained in the limit $\omega \tau \ll 1$. Given that $\sigma \sim 10^{-3}$, the above estimate confirms that the signal from the orbital perturbation dominates the Shapiro and Einstein delay signals. 

\section{Numerical Simulation}\label{app:Numerical_Simulation}

In order to validate our analytic results, we created a full numerical simulation of the solar system in the presence of ULDM. Below, we first describe the simulation of the planetary dynamics alone, before introducing an additional acceleration due to ULDM. Some results are presented at the end of this appendix. 

\subsection{Solar System Dynamics}

The simulation accounts for the major solar system objects: the sun, the eight planets from Mercury to Neptune, and the Earth's moon. For simplicity, we consider each planetary system (so, for example, Saturn and all its moons) as a single object, with the exception of Earth, where the moon is modeled separately. All objects $i$ are assumed to be point masses which are characterized by their position $\mathbf{x}_i$ and velocity $\mathbf{v}_i$ with respect to the solar system's center of gravity. The masses of all objects, as well as their initial state position and velocities, are taken from the DE-441 ephemeris model~\cite{2021AJ....161..105P} as implemented in NASA's \texttt{spiceypy} library~\cite{Annex2020, ACTON199665, ACTON20189}.

In Newtonian mechanics, the acceleration $\mathbf{a}_i$ of a solar system object is given by  
\begin{equation}
\mathbf{a}^\text{Newton}_i = \sum_{j\neq i} \frac{G M_j }{|\mathbf{x}_{i}-\mathbf{x}_j|^3} (\mathbf{x}_{j}-\mathbf{x}_i) \;, 
\label{a_Newton}
\end{equation}
where we summed over all other objects $j \neq i$ with mass $M_j$.  For sufficiently small separations, general relativistic effects can play a role. Let us define the Schwarzschild radius of an object with mass $M$ as $R_s = 2 G M / c^2$, the separation of two objects as $\mathbf{r} = \mathbf{x}_j - \mathbf{x}_i$, and their relative velocity as $\mathbf{v} = \mathbf{v}_j - \mathbf{v}_i$. If $r = |\mathbf{r}|<10^{9} R_s$, we take into account the first post-Newtonian (1PN) correction 
\begin{equation}
\mathbf{a}^\text{1PN}
= - \frac{GM}{r^3}
\left[
\left(2\frac{R_s}{r} - \frac{v^2}{c^2}\right)\mathbf{r}
+ \frac{4(\mathbf{r}\cdot\mathbf{v})}{c^2}\mathbf{v} 
\right] \; . 
\label{1PN}
\end{equation}

We can then describe the entire system with the state vector $\mathbf{X} = ( \mathbf{x}_1, \mathbf{v}_1, \mathbf{x}_2, \mathbf{v_2}, ... ).$ The equation of motion are then given by $d\mathbf{X}/dt = f(\mathbf{X},t)$ with $ f(\mathbf{X},t) = ( \mathbf{v}_1, \mathbf{a}_1,  \mathbf{v}_2, \mathbf{a}_2, ... ) $. We numerically solve this differential equation using the Runge-Kutta-4 method. Since we are interested in sub-mm precision for positions at multi-AU scales, all relevant variables are described by floating numbers with 30 digits as provided by the \texttt{boost.multiprecision} package. 

We have validated our simulation by comparing the obtained trajectories to those provided by the \texttt{Python} wrapper \texttt{spiceypy} for NASA's  library. An accumulated difference of about 1~km in the planetary position after a time span of one year was observed. This difference results from the limited cm-scale numerical precision of the provided initial state coordinates. Additional solar system objects, such as dwarf planets and asteroids, were found to have an even smaller impact. The difference between two simulations with the same initial state was at the $\mu$m level, and can be considered negligible. We have further tested the simulation by injecting a primordial black hole, as considered in Fig.~3 of Ref.~\cite{Tran:2023jci}, and were able to reproduce their result. 

\subsection{Ultralight Dark Matter}

Let us now consider the ULDM field $\phi(\mathbf{r},t)$ with mass $m$ and velocity dispersion $\sigma$ as introduced at the beginning of Sec.~\ref{sec:ULDM}. Let us further introduce the corresponding wavelength as $\lambda_{\rm DM} = 1/m\sigma$ and momentum dispersion as $\sigma_k = m \sigma$. 

We numerically compute the density across a three-dimensional grid of size $N^3$ and dimension $V=L^3$. The dimensions are determined such that it is larger than the solar system, $L > 100$~AU; significantly larger than the dark matter wavelength, $L > 10 \lambda$; and sufficiently large such that the period of the slowest mode is $T_{min} = 2\pi/\omega_{\text{min}}= m L^2/\pi$ is larger than the simulation time $T$, so $L > (\pi T / m)^{1/2}$. Furthermore, the grid spacing $L/N$ is required to be smaller than a fraction of the wavelength, $L/N < \lambda_{\rm DM}/5$. The points of the position and momentum grid are given by $x_{j} = j \cdot L/N$ and $k_{j} = j \cdot 2\pi / L$ for $j \in [-N/2, ... N/2]$ for each of the three dimensions.

The ULDM field $\phi$ can be expanded in terms of the wavefunction $\psi$ as
\begin{equation}
\phi(\mathbf{x},t) = (2 m)^{-1/2}   [\psi(\mathbf{x},t) e^{-imt} + h.c]\;, 
\end{equation}
where we have factored out the fast oscillating mode. The wavefunction $\psi$ evolves according to the Schrödinger equation. In the absence of a potential, the wavefunction can be modeled as the sum of many freely propagating plane waves
\begin{equation}
\psi(\mathbf{x},t) = V^{-1/2}  \sum_j r_j e^{i \theta_j} e^{i (\mathbf{k}_j \cdot  \mathbf{x} - \omega_j t)} \;,
\end{equation}
where the sum over $j$ runs over the momenta of a particle quantized in a cubic box of volume $V$ and $\omega_j = \mathbf{k}_j^2/2m$. The coefficients $r_j$ and $\theta_j$ describe the initial state of the dark matter distribution and are sampled according to their probability density functions given in Eq.~\eqref{eq:pdf_random}. This uses the  momentum distribution $f_j$ of the dark matter halo, which is approximately given by a normal distribution with dispersion $\sigma_k$ 
\begin{equation}
f_{j} = \frac{\bar \rho}{m}  \frac{(2\pi)^{3/2}}{\sigma_k^3} e^{-\mathbf{k}_j^2/2\sigma_k^2} \; .  
\end{equation}
In practice, the simulation is performed in dimensionless units  
\begin{align}\label{}
  \tilde{\mathbf{x}}_j &= \mathbf{x}_j \cdot \sigma_k\;,       
  \\
  \tilde{\mathbf{k}}_j &= \mathbf{k}_j/\sigma_k \;, 
\\
\tilde t &= t \cdot \sigma_k^2 / m\;, 
\\
\tilde \omega_k &= \omega_{k} \cdot m / \sigma_k^2 \;,
\\
\tilde f_j&= f_j \, / \, (\bar \rho / m ) \;, 
\\
\tilde r_j &= r_j \, / \, \sqrt{\bar \rho / m } \;, 
\\
\tilde \psi &= \psi \, / \, \sqrt{\bar \rho / m } \; . 
\end{align}
In this dimensionless unit, length and momentum are measured in wavelength and momentum dispersion of dark matter, and time and energy are measured in coherence time and typical kinetic energy of dark matter.

The force on each planet due to ultralight dark matter is directly computed from the Fourier component of the gravitational potential,
\begin{align}
\mathbf{a}_i^\text{ DM}(\mathbf{x}_i)
= - \nabla \Phi(\mathbf{x}_i)
= - \int \frac{d^3 k}{(2\pi)^3} e^{i \mathbf{k} \cdot \mathbf{x}_i} i \mathbf{k} \tilde \Phi(\mathbf{k}) , 
\label{a_DM}
\end{align}
where $\mathbf{x}_i$ is the position of the planet $i$. To compute the Fourier coefficient of the gravitational potential, we use the Possion equation, $\nabla^2 \Phi = 4 \pi G \delta\rho$, from which we obtain
\begin{align}
\tilde \Phi({\mathbf k}) = - \frac{4\pi G}{k^2} \delta\tilde \rho(\mathbf{k}) . 
\end{align}
The density fluctuation is obtained by subtracting the mean density from the density, $\delta\rho(\mathbf{x}) = m|\psi(\mathbf{x})|^2 - m \langle |\psi(\mathbf{x})|^2\rangle$, where the angle bracket represents an averaging operation over the entire simulation box. The total force experienced by body $i$ is then obtained by summing the force due to other solar system objects, Eqs.~\eqref{a_Newton}--\eqref{1PN}, and the above dark matter force, Eq.~\eqref{a_DM}.

\subsection{Simulation Results}
\begin{figure*}[!t]
\centering
\includegraphics[width=0.328\linewidth]{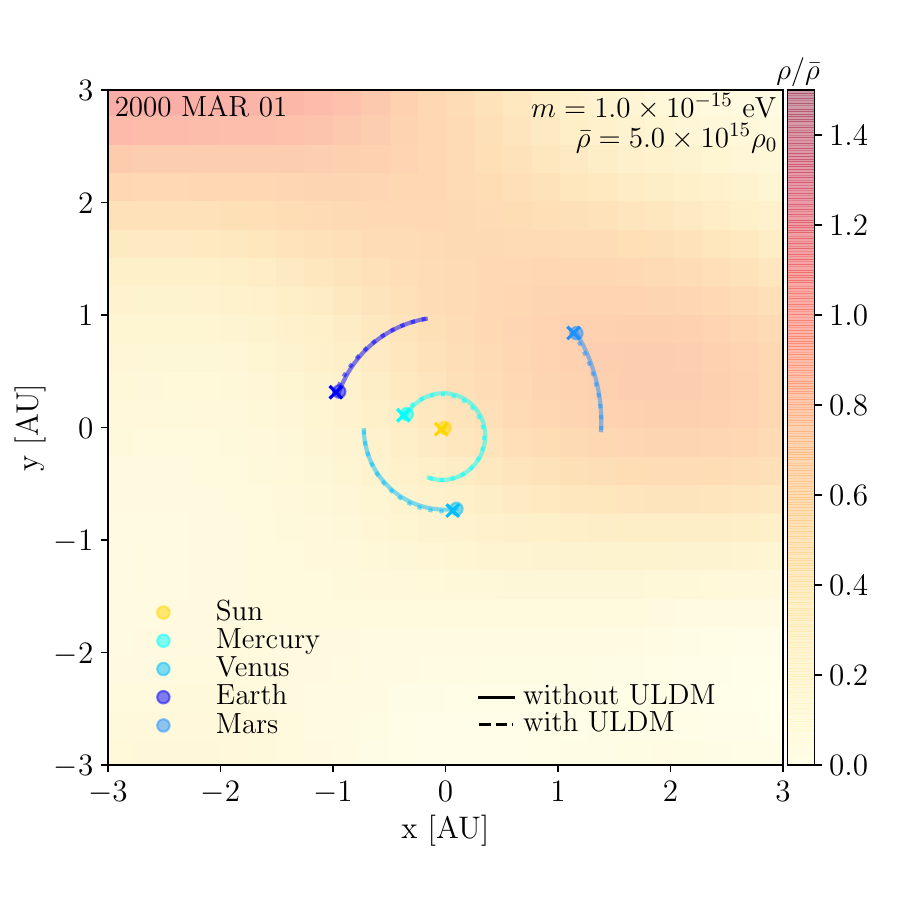}
\includegraphics[width=0.328\linewidth]{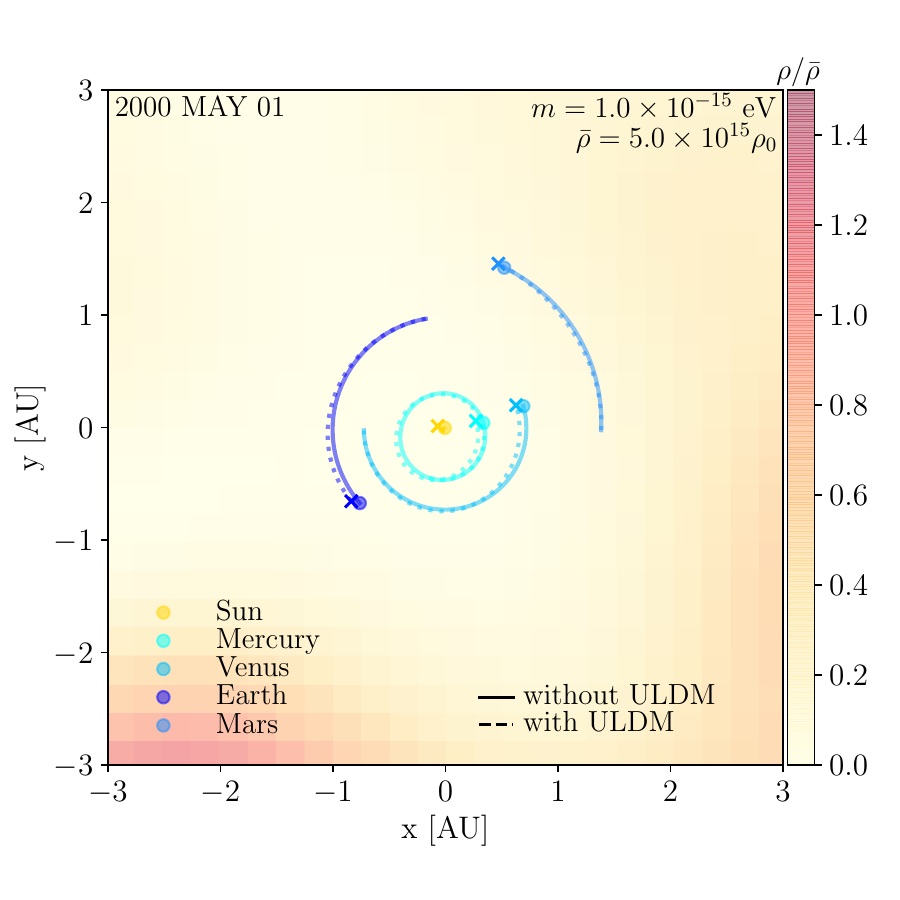}
\includegraphics[width=0.328\linewidth]{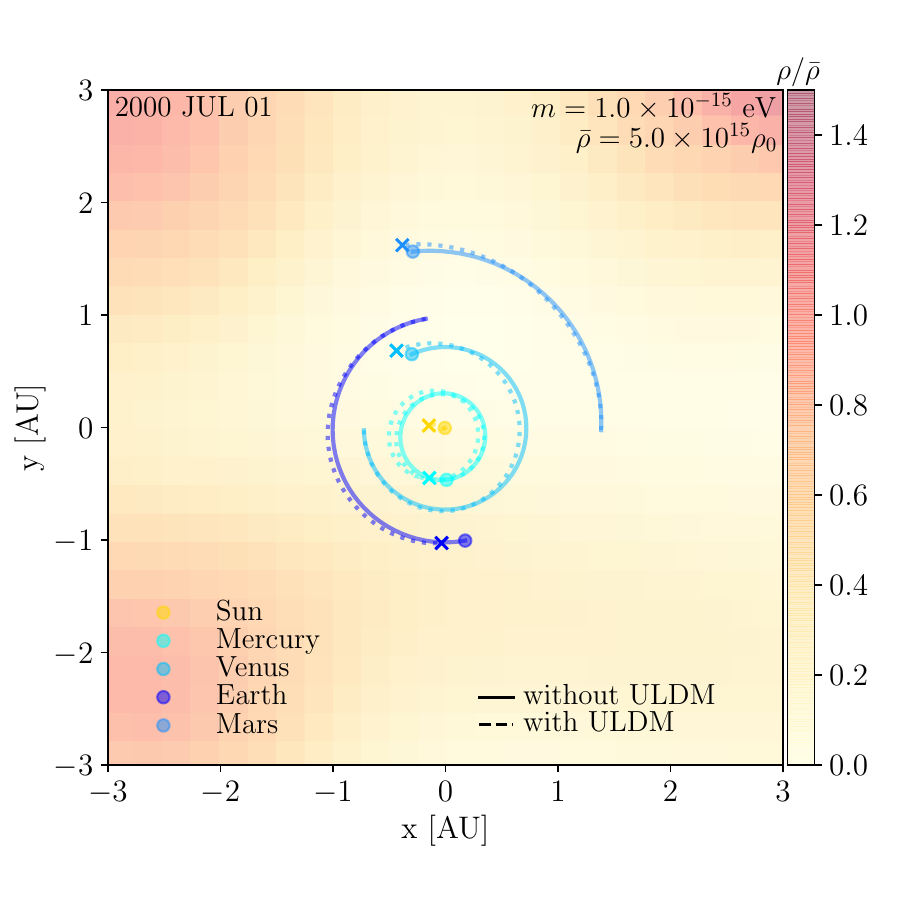}
\caption{Example of simulation results. We simulate the planetary trajectories in a simplified model of the solar system, consisting only of the sun and inner planets, in the absence and presence of ULDM. We consider an ULDM mass of $m=10^{-15}$~eV. To obtain visible deviations of the trajectories on AU scales, an average dark matter density $\bar \rho = 5\cdot 10^{15} \rho_0$ was chosen. The simulation starts on Jan 1st 2000, with the planets being at their nominal positions. The figures show snapshots of the dark matter density in the planetary plane and the planetary trajectories in the absence (solid) and presence (dashed) of ULDM after two, four, and six months.}
\label{fig:simulation_example}
\end{figure*}

An example of planetary trajectories in the presence of ULDM obtained using the numerical simulation is shown in Fig.~\ref{fig:simulation_example}. Here we consider a ULDM candidate with mass $m=10^{-15}$~eV and average density $ \bar{\rho} = 5\cdot 10^{15} \rho_0$, and coefficients describing the initial state sampled according to their distributions. The color scale in the background shows the ULDM density $\rho/\bar{\rho}$ in the $z=0$ plane at three different dates. We can clearly see that the appearance of under- and over-densities with spatial extension $L \sim (\sigma m)^{-1} \sim 1$~AU. Superimposed are the trajectories of the inner planets in the absence (solid curves) and presence (dashed curves) of ULDM. We can see that the planets visibly deviate from their nominal trajectory in the presence of ULDM. 

\begin{figure*}[!t]
    \centering
    \includegraphics[width=0.49\linewidth]{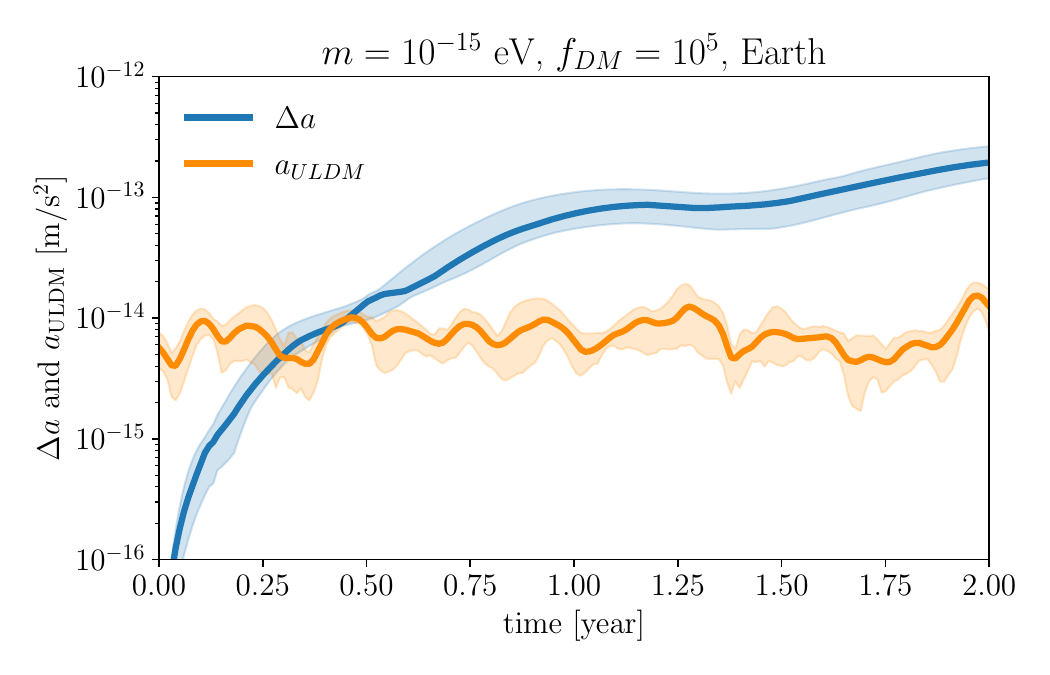}
    \includegraphics[width=0.49\linewidth]{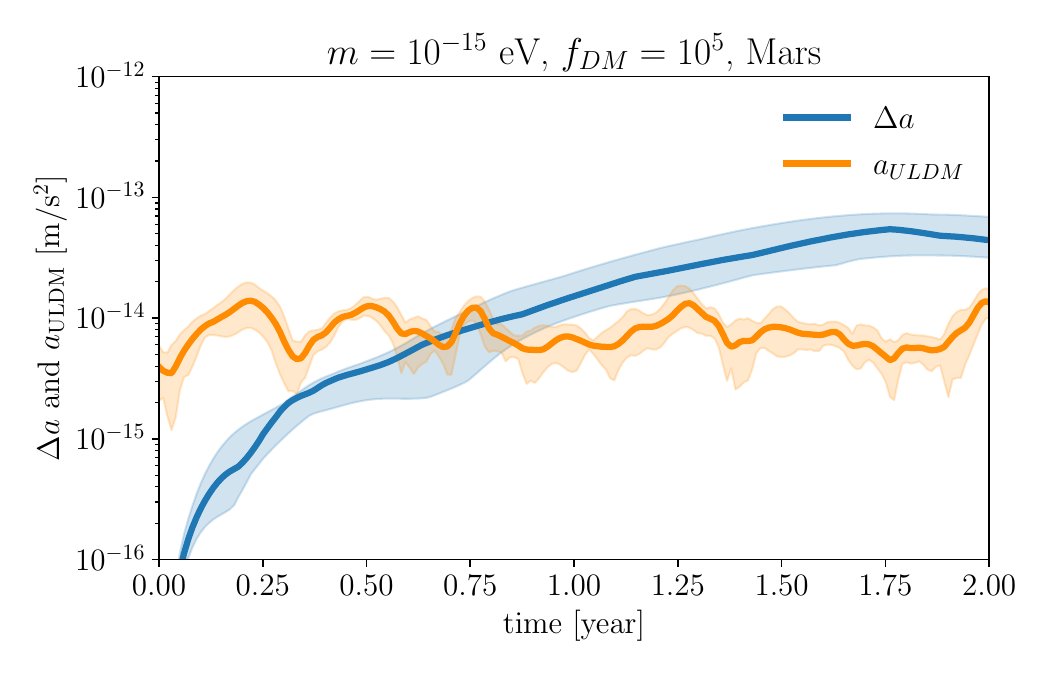}
    \includegraphics[width=0.49\linewidth]{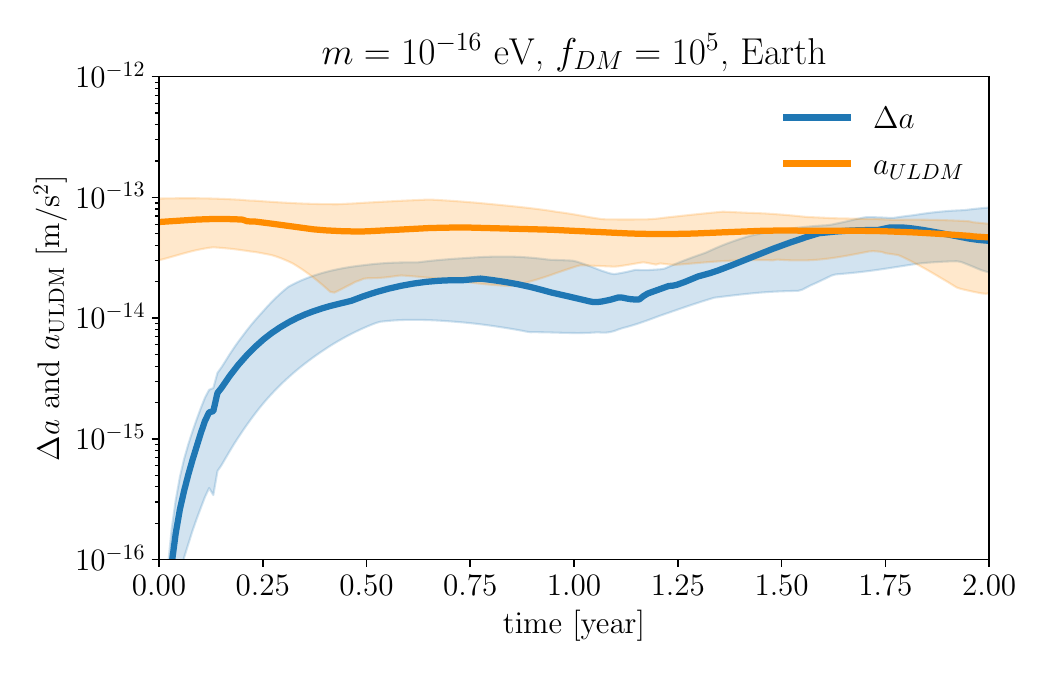}
    \includegraphics[width=0.49\linewidth]{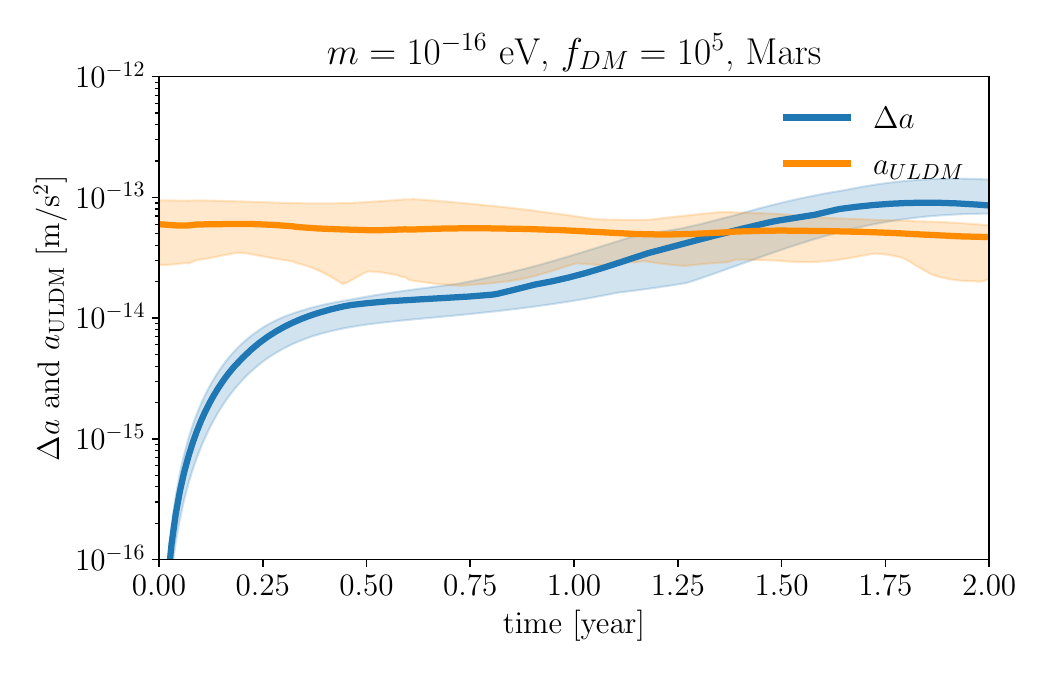}
    \caption{Acceleration experienced by Earth (left) and Mars (right) due to ULDM, $a_\text{ULDM}$, and the difference of their acceleration due the Sun and other planets between the nominal trajectory (without ULDM) and perturbed trajectory (with ULDM), $\Delta a$. 10 ULDM simulations with different random initial states were performed for $m=10^{-15}$~eV (top) and $m=10^{-16}$~eV (bottom). The figure shows the average accelerations and the corresponding uncertainty band. The difference in acceleration that Mars and Earth experience on their perturbed trajectory compared to their nominal trajectory exceeds the acceleration due to dark matter after about half a year in all simulations. This indicates that non-linear effects become dominant at longer time scales.}
    \label{fig:acc_feedback}
\end{figure*}

When the planets deviate sizably from their nominal trajectory, the gravitational force from the other celestial objects acting on them also changes. If sufficiently large, this difference in the gravitational force will cause the spatial separation of the planets from their nominal orbit to evolve non-linearly. Let us study this effect by considering the difference of acceleration $\Delta a$ of a given planet due to all other solar system objects between its perturbed trajectory and nominal trajectory. The blue curves in Fig.~\ref{fig:acc_feedback} show $\Delta a$ for Earth and Mars as a function of time for two benchmark scenarios with ULDM masses $m=10^{15}$~eV and $10^{-16}$~eV and average ULDM density $\bar\rho = 10^5 \rho_0$. The solid line and shaded band show the average prediction and standard deviation, respectively, which were obtained from 10 simulations with different randomly sampled initial states. 

For comparison, we show the gravitational acceleration directly due to the ULDM $a_\text{ULDM}$ as an orange line, again obtained from 10 separate simulations. We see that $\Delta a > a_\text{ULDM}$, after about half a year for $m=10^{-15}$~eV and about 1.5~years for $m=10^{-16}$~eV. This indicates that linear effects are dominant on short timescales while non-linear effects accumulate over a time scale of the orbital period. Using the simulation, we also find that for the inner planets, Venus and Mercury, non-linear effects accumulate over a shorter time scale due to their shorter orbital period, while for the outer planets beyond Jupiter, they become important significantly later. 

\begin{figure*}
    \centering
    \includegraphics[width=0.43\linewidth]{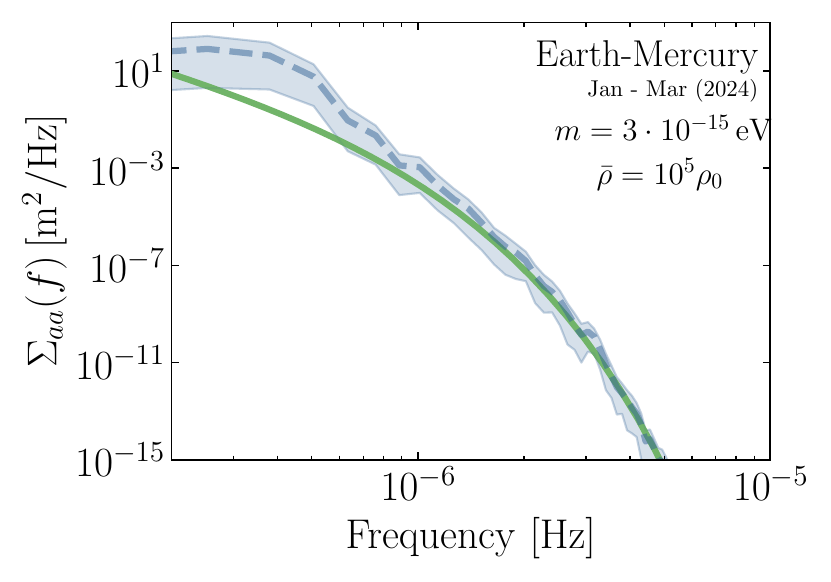}
    \includegraphics[width=0.43\linewidth]{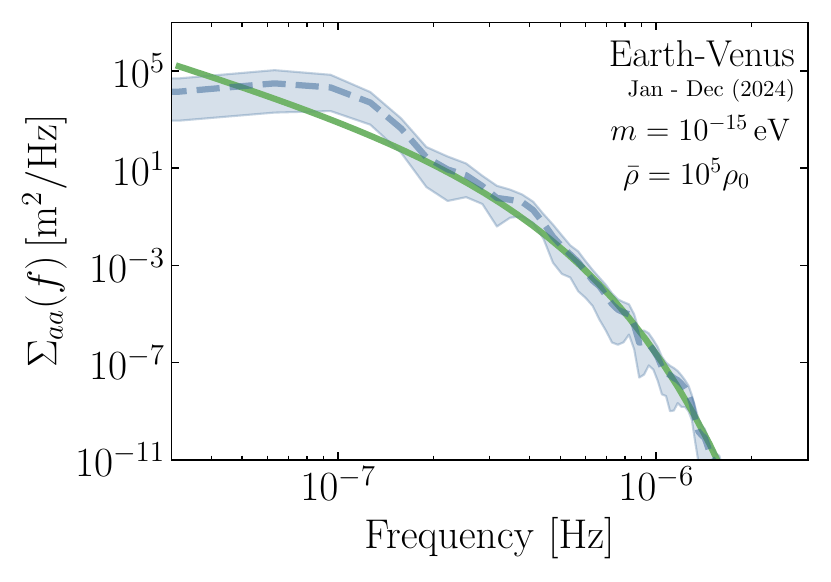}
    \\
    \includegraphics[width=0.43\linewidth]{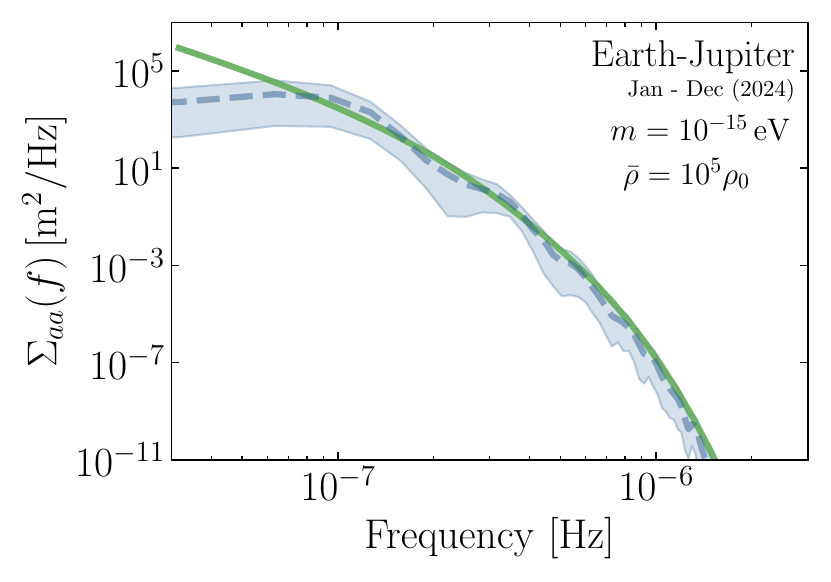}
    \includegraphics[width=0.43\linewidth]{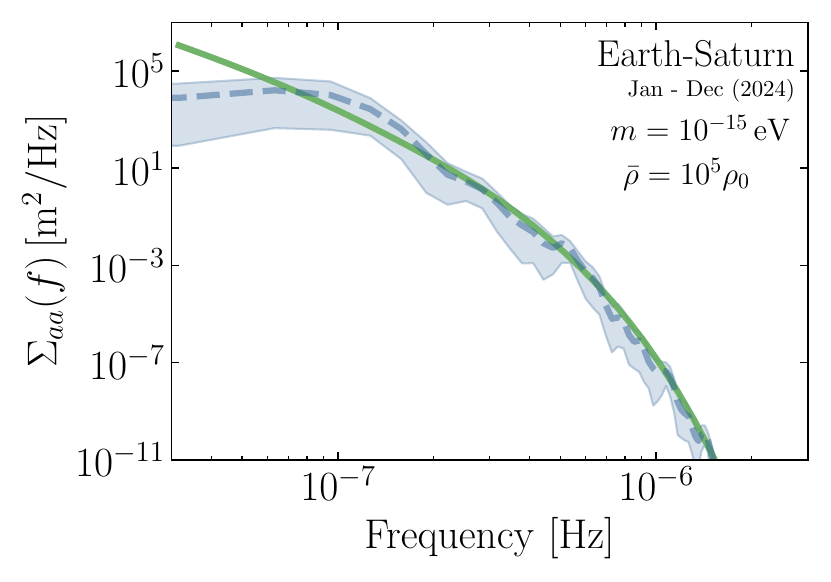}
    \caption{Numerically estimated auto-correlation for each residual range. We performed 100 numerical simulations. The dashed line is the mean of the power spectral estimated from each numerical simulation, and the upper and lower boundary are the $5\%$ and $95\%$ quantiles, respectively. Except for Earth-Mercury, all results are based on simulations over the entire year 2024. The Earth-Mercury result is obtained from numerical simulations during January 2024 -- March 2024 with a different mass $m=3\times 10^{-15}\eV$.}
    \label{fig:psd_simulation}
\end{figure*}

In Fig.~\ref{fig:psd_simulation}, we show the reconstruction of the signal power spectrum for each residual range. For these plots, we performed $100$ independent simulations in the year $2024$. We follow the procedure described in the main text; we apply the window function with $\alpha=6$ to the residual range, detrend the series with a polynomial of degree $n=4$, and construct the power spectrum from the discrete Fourier transformation of the resulting time series. The results for Earth-Venus, Earth-Jupiter, and Earth-Saturn agree with the analytic estimation at the relevant frequencies. 

The case of Earth-Mercury deserves particular attention. As discussed in the main text, our analytic estimation is obtained over the time scale at which the nominal separation remains approximately constant. Since the orbital period of Mercury is about three months, we do not expect that the analytic result agrees with the numerically constructed power spectrum. In fact, we observed that under the same simulation conditions---100 independent simulations over the year 2024---we were unable to reconstruct numerically the power spectrum expected from our analytic formulae. Only when we restricted the time series to be shorter than the orbital period of Mercury were we able to reconstruct the power spectrum that agrees reasonably with the analytic result. Specifically, for the Earth-Mercury range, we performed a similar type of 100 numerical simulations but only over January 2024 to March 2024. We also choose a slightly heavier dark matter mass $m=3\times 10^{-15}\eV$ for demonstration. The result is shown in the upper left panel of Fig.~\ref{fig:psd_simulation}.

\section{Systematic Estimation of Cross-Correlation}\label{app:systematic}
In this section, we provide a systematic investigation of the correlation coefficients for the free particles and provide a detailed discussion of how we estimate the correlation coefficient of range measurements presented in the main text. The main purpose of this discussion is to validate the analytic result on the correlation coefficients as a function of angle with numerical simulations. 

\subsection{Free Particles}
We consider a simple scenario of three free particles. One of them represents the Earth, while the other two represent different planets. We refer to the one representing Earth as the reference particle, and the others representing a different planet as planet $a$ and $b$, respectively. We assume that all particles experience force only due to ultralight dark matter. In this setup, the subtended angle remains constant in the absence of ultralight dark matter. This setup enables a systematic study of the correlation coefficient $\Gamma_{ab}$ as a function of the subtended angle $\hat{\boldsymbol n}_a \cdot \hat{\boldsymbol n}_b$, frequency, the nominal separation $\bar L_{a,b}$, and the mass of dark matter. Here $\hat{\boldsymbol n}_{a,b}$ is the unit vector from the reference to the planet $a$ and $b$. 

We perform numerical simulations by initializing three particles in a constellation defined by the two arm lengths $\bar L_{a,b}$ and the angle $\hat{\boldsymbol n_a} \cdot \hat{\boldsymbol n}_b = \cos\theta$. We simulate the system in a square box, whose side is given by $L=12$ with $N=64$ grid points. Here, we use the dimensionless grid units established in the previous section. As discussed above, the size of the simulation sets the maximum time above which we begin to see periodicity of the dark matter evolution. This time is given by $T=L^2/\pi\simeq 45.8$. We choose the number of time steps as $N_t = N^2 = 4096$. From each simulation, we obtain two time series of the fluctuations in the separation of planets with respect to the reference particle, $\Delta L_{a,b}(t)$.

We compute the power spectral density for each residual range and cross-spectral density. To estimate them, we utilize the Welch method as implemented in \texttt{SciPy}~\cite{2020SciPy-NMeth}. In particular, we use \texttt{scipy.signal.welch} for the power spectral densities and \texttt{scipy.signal.csd} for the cross-spectral density. The idea is to dissect the time series of duration $T$ into segments of duration $T_\text{seg}$. Each segment is applied to a window function and discrete Fourier-transformed to estimate the power spectral density. Neighboring segments are partially overlapped to compensate for the loss of information from the application of the window function. In our time series of length $N_t=4096$, we choose the segment length to be $N_{\rm seg}= 1024$ with a window function (cf.\ Eq.~\eqref{window}) with $\alpha=4$ to suppress spectral leakage. 

We select a few benchmark points at which we check the numerical result against the analytical results. In particular, we choose different arm lengths $\bar L_{a}= \bar L_b\in \{0.5, \, 1, \, 5\}$, different frequencies $f \in \{ 0.1,  \, 0.5, \, 1 \}$, and  angles $\theta \in [0,\pi]$. For a given set of parameters, we perform $N_{\rm sim}=100$ simulations for the statistical evaluation of the correlation coefficient. 

\begin{figure*}[ht!]
\centering
\includegraphics[width=0.32\textwidth]{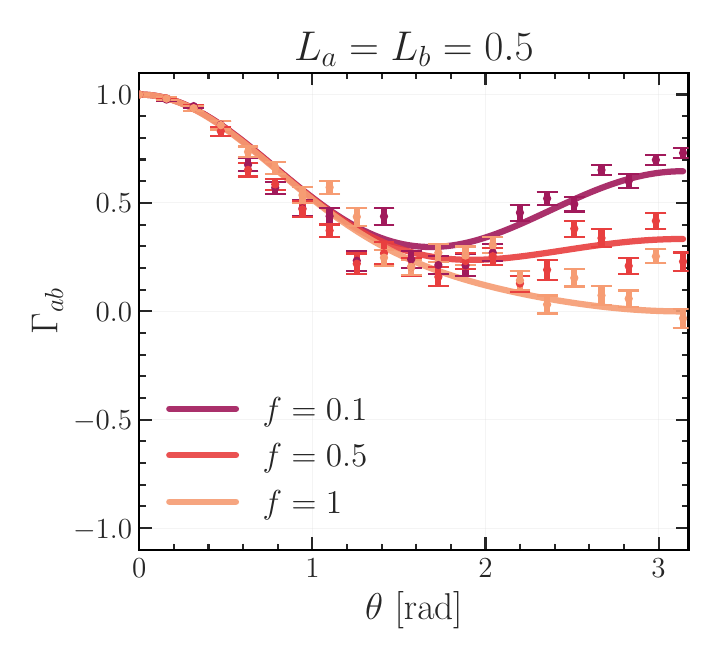}
\includegraphics[width=0.32\textwidth]{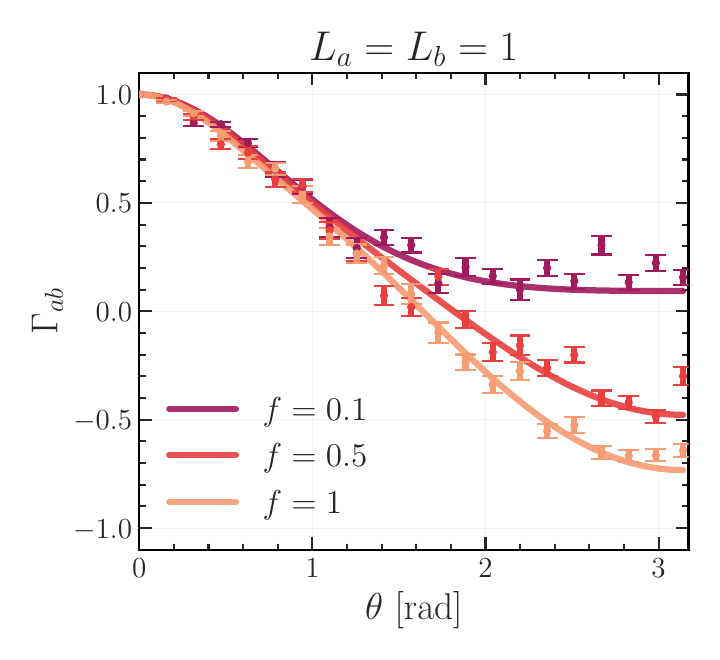}
\includegraphics[width=0.32\textwidth]{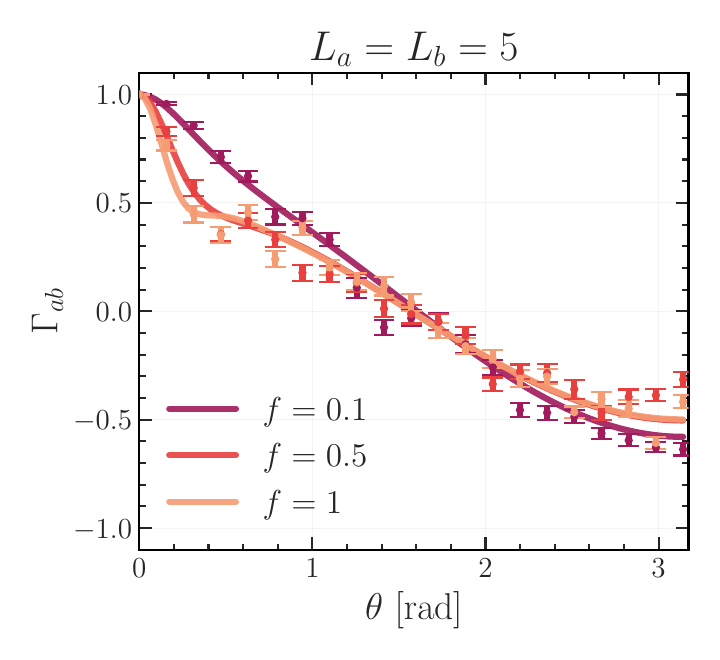}
\caption{The correlation coefficients. The bars and lines are numerical and analytical results, respectively. The size of the bar is the error of the estimator given in the text. For all benchmark points we tested, the numerical result agrees with the analytic estimation. See the text for details.}
\label{fig:orf_dimless}
\end{figure*}

The result is shown in Fig.~\ref{fig:orf_dimless}. What we compute is the following averaged correlation coefficient:
\begin{align}
\hat \Gamma_{ab} = \frac{1}{N_{\rm sim}} \sum_{i=1}^{N_{\rm sim}} \hat \Gamma_{ab}^{(i)}
\end{align}
where $\hat \Gamma_{ab}^{(i)}$ is the correlation coefficient estimator from a single simulation. The central point and the error bar denote the mean and the standard deviation of the above estimator. Assuming that all simulations are fully independent, we estimate the latter as the standard deviation of the $N_\text{sim}$ simulations divided by $\sqrt{N_\text{sim}}$ instead of doing repeated measurements of  $\hat \Gamma_{ab}$. The solid lines are the analytic results. The plot shows good agreement between numerical and analytical results for all choices of the separation, frequency, and dark matter mass we considered. Note that the analytic result is obtained from the one-way response and that we assume $\omega \bar L \ll 1$ as in the physical case discussed in the main text. 

It is instructive to convert the result of the dimensionless simulation to the physical one. The length unit is given by the wavelength of dark matter. If we assume that the wavelength is given by $1\,{\rm AU}$, this corresponds to the dark matter mass $m =2\times 10^{-15}\,{\rm eV}$ for the standard halo model with $\sigma=160\,{\rm km/sec}$. The duration of the simulation is given by $1.4\,{\rm yr}$, while the simulation time step is about $1.2$ hours. 

\subsection{Orbiting Particles}
The systematic investigation of the correlation coefficient as a function of angle is cumbersome when the particles are orbiting around the Sun. This is because the subtended angle constantly changes in time. For this reason, in the main text, we present the correlation coefficient as a function of time, not as a function of angle. 

As the numerical computation of the correlation coefficient of orbital particles---the one given in Fig.~\ref{fig:spectrum_time}---involves a few subtleties, we provide a more detailed discussion of the procedure for the estimation. As in the case of free particles, we begin by numerically estimating the power spectral density as well as the cross-spectral density. The difference to the free particle case is that the signal power spectrum can easily be dominated by the power due to the effective restoring force (cf.\ sec.~\ref{sec:sol_acc}) at the orbital frequency of the planets, if the orbital motion is not subtracted from the data stream before the numerical estimation of the spectral densities. For this reason, we subtract the separation between planets with a so-called nominal separation obtained without dark matter, and we further detrend the time series with a polynomial of degree $n=4$. With this detrending, we aim to subtract the effect of a slow drift of orbital trajectories away from the nominal one due to dark matter induced perturbation. 

This procedure slightly distorts the signal power spectrum at the lower end of available frequencies. This is because, although it eliminates the effect of the orbital motion to a sufficient degree, it also partially removes the dark matter signal. At the high frequency tail, our procedure is effective at reconstructing the analytical prediction, as discussed before. Nevertheless, the signal is exponentially suppressed at $\omega > m \sigma^2$, making it increasingly difficult to infer from the simulation.

For this reason, we chose a frequency around $\omega \sim m\sigma^2$ where we can reconstruct the signal power spectrum with the greatest precision. We find $f= 7\times 10^{-6}\,{\rm Hz}$ for $m=10^{-14}\,$eV is most suitable for this demonstration. Practically, to derive the correlation coefficient, we run $N_{\rm sim}=100$ simulations in each year from 2024 to 2027. The initial conditions of each simulation are fixed by the planetary ephemeris data provided by \texttt{spiceypy} library. For each simulation, we segment the time series into 6 intervals, which we analyze using the Welch method with four subsegments. From this analysis, we obtain estimates of the correlation coefficient at 6 different points in a given year. The result of the $N_{\rm sim}=100$ simulations forms a distribution of these estimates, which is shown as violins in Fig.~\ref{fig:spectrum_time}.

\bibliographystyle{utphys}
\bibliography{ref}

\end{document}